\documentclass[12pt]{iopart}
\usepackage{iopams}
\usepackage{graphicx}

\def\rt{\mathfrak{r}}
\def\at{\mathfrak{a}}
\def\bt{\mathfrak{b}}
\def\bV{\mathrm{V}}
\def\bW{\mathrm{W}}
\newtheorem{theo}{Theorem}
\newtheorem{defi}{Definition}
\newtheorem{prop}{Proposition}
\begin{document}
\title{Large $N$ expansions and Painlev\'e hierarchies in the Hermitian matrix model}
\author{ Gabriel \'Alvarez$^1$, Luis Mart\'{\i}nez Alonso$^1$ and Elena Medina$^2$}
\address{$^1$ Departamento de F\'{\i}sica Te\'orica II,
                        Facultad de Ciencias F\'{\i}sicas,
                        Universidad Complutense,
                        28040 Madrid, Spain}
\address{$^2$ Departamento de Matem\'aticas,
                        Facultad de Ciencias,
                        Universidad de C\'adiz,
                        11510 Puerto Real, C\'adiz, Spain}
\begin{abstract}
We present a method to characterize and compute the large $N$ formal asymptotics of regular and critical
Hermitian matrix models with general even potentials in the one-cut and two-cut cases. Our analysis is based
on a method to solve continuum limits  of the discrete string equation which uses the resolvent of the Lax operator of the underlying
Toda hierarchy. This method also leads to an explicit formulation, in terms of coupling constants and critical parameters,
of the members of the Painlev\'e~I and Painlev\'e~II hierarchies associated with one-cut and two-cut critical models respectively.
\end{abstract}
\submitto{\JPA}
\maketitle
\section{Introduction}
The study of the  asymptotic properties of random matrix models in the limit of large matrix size has revealed deep connections
between random matrix theory and the theory of integrable systems. The origin of these connections is nowadays much better
understood thanks to the use of Riemann-Hilbert (RH) techniques, which are one of the main tools in the inverse scattering theory
of integrable systems~\cite{FO91,FO92, DE99,DE99b,BL08}. Indeed, the occurrence of RH problems leads naturally to Lax pair
equations and to the corresponding string equations~\cite{MA09}, which in turn  explain why integrable equations of Painlev\'e type
arise in the double scaling limit of critical random matrix models~\cite{DI95}. The main goal  of this work is to describe in detail
and in a general setting (i.e. independently of particular examples which will be used only for illustrative purposes)
the correspondence between the full Painlev\'e~I and Painlev\'e~II hierarchies  and critical random matrix models. 

Specifically, we consider  random Hermitian matrix models with partition function
\begin{equation}
  \label{0.1}
  Z_{N} = \int_{\mathbf{R}^N}
                \rme^{-N  \sum_{i=1}^N V(x_i)}
                \prod_{i<j} (x_i-x_j)^2
                \rmd x_1 \cdots \rmd x_N,
\end{equation}
where the integration is performed over the eigenvalues $x_1,\ldots,x_N$ and $V$ is a polynomial potential of even degree
and real coefficients (coupling constants) $\mathbf{g}=(g_1,\ldots,g_{2p})$,
\begin{equation}
  V(z) = \sum_{n=1}^{2p} g_n z^{n},\quad g_{2p}> 0.
\end{equation}
The  large $N$ asymptotics of a matrix model is related to the large $N$ asymptotics of the orthogonal polynomials 
$P_{n,N}(x)$ with respect to the exponential weight $\rme^{-N V(x)}$, i.e. monic polynomials
\begin{equation}
  \label{mon}
  P_{n,N}(x) = x^n + a_{n-1} x^{n-1} + \cdots
\end{equation}
that satisfy
\begin{equation}
  \label{pol}
  \int_{-\infty}^{\infty} P_{k,N}(x) P_{l,N}(x) \rme^{-N V(x)} \rmd x = \delta_{kl} h_{k,N}.
\end{equation}
In fact there are several methods~\cite{BR78,BE79,BE80,ER03,BL05} that reduce the calculation of the large $N$
asymptotic behavior of the free energy
\begin{equation}
  F_N = -N^{-2} \ln Z_N
\end{equation}
to the calculation of the asymptotic behavior of the recurrence coefficients $r_{n,N}$ and $s_{n,N}$ in the three-term
recursion relation
\begin{equation}
  \label{rec}
  x P_{n,N}(x) = P_{n+1,N}(x) + s_{n,N} P_{n,N}(x) + r_{n,N} P_{n-1,N}(x).
\end{equation}

The structure of these asymptotic expansions depends on the number of
disjoint intervals (cuts) in the support $J$ of the asymptotic eigenvalue density $\rho(x)$.
In fact, the asymptotic expansions in the multi-cut (two or more cuts) case are not in general simple power
series in $N^{-1}$ but have a complicated quasi-periodic dependence on $N$~\cite{DE99b,BL08,BO00,EY09}.
However, for even potentials the recurrence coefficient $s_{n,N}$ is identically zero, and if in addition the
potential is in the regular (noncritical) two-cut case with eigenvalue support $J=(-\beta,-\alpha)\cup (\alpha,\beta)$,
then a substantial simplification of the general behavior occurs~\cite{BO00}: the odd terms of the recurrence
coefficient $r_{2n+1,N}$ and the even terms $r_{2n,N}$ admit (different) asymptotic expansions
whose leading terms can be gathered in a single formula,
\begin{equation}
  \label{asy1}
  r_{n,N} = \frac{1}{4} \left(\alpha-(-1)^n \beta\right)^2 + \Or(N^{-2}),\quad N\rightarrow \infty.
\end{equation}

Our goal is to study the critical behavior of these matrix models with respect to a temperature parameter $T>0$,
i.e.~we consider the family of models $Z_N(T)$ with potentials $V(z)/T$ or, equivalently, with coupling constants
$\mathbf{g}/T$. Critical matrix models correspond to points $T_\mathrm{c}$ where the asymptotic free energy
ceases to be analytic as a function of $T$. Some of these critical models are related to conformal $(p,q)$ minimal
models~\cite{DI95,DI97,BER09,MA10} and have asymptotic eigenvalue densities with rational singularities
\begin{equation}
  \label{den}
  \rho(x) \sim (x-\alpha)^{p/q}
\end{equation}
near one of the endpoints $\alpha$ of $J$.
It has been known for a long time in the physics literature that the double scaling limit of critical matrix models in the
one-cut and two-cut cases reveals the presence of equations belonging to the Painlev\'e~I and Painlev\'e~II hierarchies
respectively (see~\cite{KU03} for a description of these hierarchies). However, most of the existing work,
specially for the two-cut case, deals only with particular
examples~\cite{BL08,DI95,BL05,DE90,BL99,BL03b,BL03} and does not offer a general characterization
of the specific member of the Painlev\'e hierarchy associated to a critical model in terms of the coupling constants
$\mathbf{g}$ and of the critical value of the recurrence coefficient. More recent works~\cite{BER09,MA10}
do prove the occurrence of the full Painlev\'e hierarchies in the context of critical matrix models, but again they do not
give an explicit correspondence between Painlev\'e equations and critical models.

To arrive at this general correspondence, we first perform a detailed analysis of the large
$N$ expansions of the recurrence coefficient $r_{n,N}$ for general even potentials in the one-cut
and two-cut cases, with special emphasis on the latter. Our analysis is based on a method for solving
continuum limits of the discrete string equation~\cite{DI95}. This method uses the resolvent of the Lax operator of the underlying
Toda hierarchy and can be applied to obtain the large $N$ expansions both in regular and in critical
models~\cite{MA07,MAE08}. The asymptotic behavior of $r_{n,N}$ for regular models is given
by power series in $N^{-2}$ whose coefficients can be determined recursively. For critical models we apply
the double scaling method to calculate a family of asymptotic solutions of the string equation (symmetric solutions)
which are series in fractional powers of $N^{-1}$ whose coefficients are constrained by ordinary differential
equations in the scaling variable. It is at this point when we are able to give the precise connection between 
the complete Painlev\'e~I and Painlev\'e~II hierarchies and certain families of critical models.
In particular we show how the  Painlev\'e II hierarchy is connected to critical models featuring the merging of two
cuts~\cite{MA10,BL03b,BL03}. 
 
The paper is organized as follows. In section~2 we recall the basic definitions and notations for multi-cut matrix
models and their phase spaces. In section~3  we formulate the discrete string equation in terms of a generating
function and derive some important identities. Sections~4 and~5 contain a detailed description of the large $N$
asymptotics of  the recurrence coefficient $r_{n,N}$ in the regular and critical one-cut cases respectively.
The classification of critical one-cut models and the connection to the Painlev\'e~I hierarchy contained in
section~5 is relatively simple but it sets the pattern for the rather more complicated two-cut case in
sections~6,~7~and~8.  The generality of the classification of critical two-cut models and their connection to
the Painlev\'e~II hierarchy achieved in section~8 requires a somewhat complicated calculation, but the
final result can be stated concisely and is illustrated in the case of a merging of two cuts in
a quartic model. The paper ends with a brief summary and outlook section.
\section{Multicut models}
Let us consider a model~\eref{0.1} with coupling constants $\mathbf{g}$. It is well known~\cite{DE99,DE99b,BL08,AL10} 
that in the limit $N\rightarrow \infty$ the support $J$ of the density of eigenvalues $\rho(x)$   is the union of a finite number
$s$ of real intervals (cuts)
\begin{equation}
  \label{cut}
  J = \bigcup_{j=1}^s (\alpha_j,\beta_j),
  \quad \alpha_1<\beta_1<\cdots<\alpha_s<\beta_s,
\end{equation}
where $1 \le s \le p = (\deg V)/2$. The conditions that determine the actual value of $s$ among those allowed by this
bound can be stated in terms of a polynomial $h(z)$ defined in the following way: let $w(z)$ be the Riemann surface
\begin{equation}
  \label{rie}
  w(z) = \sqrt{\prod_{i=1}^{s} (z-\alpha_i)(z-\beta_i)},
\end{equation}
and let $w_1(z)$ be the branch of $w(z)$ with asymptotic behavior $w_1(z)\sim z^s$ as
$z\rightarrow\infty$; $h(z)$ is the polynomial part of the large $z$ expansion of $V'(z)/w_1(z)$, i.e.
\begin{equation}
  \label{0.3}
  \frac{V'(z)}{w_1(z)} = h(z) + \Or(z^{-1}),\quad z\rightarrow \infty.
\end{equation}
The $2 s$ endpoints $\alpha_1,\ldots,\beta_s$ in the $s$-cut case are solutions of the system of $2 s$ equations
\begin{eqnarray}
  \label{e1}
  \int_{\beta_{j}}^{\alpha_{j+1}}h(x)w_{1,+}(x)\rmd x=0,\quad j=1,\ldots,s-1,\\
  \label{e2}
  \oint_{\gamma}z^j\frac{V_z(z)}{w_1(z)}\rmd z=0,\quad j=0,\ldots,s-1,\\
  \label{e3}
  \oint_{\gamma}h(z)w_{1}(z)\rmd z=-4\pi\rmi,
\end{eqnarray}
where $\gamma$ is a large positively oriented loop around $J$. However, these equations may not be sufficient to
determine uniquely $s$ because they may have admissible solutions for several values of $s$.
In this case the additional condition $\rho(x)>0$ for $x\in J$ and the inequalities
\begin{eqnarray}
  \label{des1}
  \int_x^{\alpha_1} h(x')w_{1}(x')\rmd x'\leq 0,\quad \mbox{for $x<\alpha_1$} ,\\
  \label{des2}
  \int_{\beta_{j}}^x h(x')w_{1}(x')\rmd x'\geq 0,\quad \mbox{for $\beta_{j}<x<\alpha_{j+1},\quad j=1,\ldots, s-1$,}\\
  \label{des3}
  \int_{\beta_{s}}^x h(x')w_{1}(x')\rmd x'\geq 0,\quad \mbox{for $x>\beta_{s}$,}
\end{eqnarray}
characterize uniquely the solution of the problem. Finally, the polynomial $h(z)$ is related to the eigenvalue density by
\begin{equation}
  \label{0.2}
  \rho(x) = \frac{h(x)}{2\pi\rmi} w_{1,+}(x),\quad x\in J,
\end{equation}
where $w_{1,+}(x)$ denotes the boundary value of $w_1(z)$ on $J$ from above. A  model is said to be a \emph{regular}
if $h(x)\neq 0$ on $\bar{J}$ and the inequalities~\eref{des1}--\eref{des3} are strict. Otherwise it is called \emph{critical}.

As we said in the introduction we restrict our considerations to even potentials
\begin{equation}
  \label{eve}
  V(\lambda) = \sum_{j=1}^pg_{2j}\lambda^{j},
  \quad
  \lambda = z^2.
\end{equation}
For these models the eigenvalue support $J$ is symmetric with respect to the origin. In the one-cut case
 $J=(-\alpha,\alpha)$ and equations~\eref{e1}--\eref{e3} reduce to the single condition
\begin{equation}
  \label{q1}
  \oint_{\gamma}\frac{\rmd \lambda}{2\pi\rmi}
                             V_{\lambda}(\lambda)
                             \sqrt{\frac{\lambda}{\lambda-\alpha^2}} =1,
\end{equation}
where we have denoted $V_{\lambda}(\lambda)=\rmd V(\lambda)/\rmd\lambda$.
Likewise, in the two-cut case  $J= (-\beta,-\alpha) \cup (\alpha,\beta)$ and equations~\eref{e1}--\eref{e3} reduce to
\begin{equation}
  \label{q2}
  \eqalign{
                \oint_{\gamma}\frac{\rmd \lambda}{2\pi\rmi}
                 \frac{V_{\lambda}(\lambda)}{\sqrt{(\lambda-\alpha^2)(\lambda-\beta^2)}} =0,\\
                 \oint_{\gamma}\frac{\rmd \lambda}{2\pi\rmi}
                 \frac{\lambda V_{\lambda}(\lambda)}{\sqrt{(\lambda-\alpha^2)(\lambda-\beta^2)}}=1.}
\end{equation}

Given a family of potentials~\eref{eve} with the  coupling constants $\mathbf{g}=(g_2,g_4,\ldots,g_{2p})$
running on a certain region $G$ of $\mathbf{R}^p$, we decompose the region $G$ as
\begin{equation}
  \label{cou}
  G = \bigcup_{s\geq 1}\overline{G}_s,
\end{equation}
where $\mathbf{g}\in G_s$ if and only if $\mathbf{g}$  determines a $s$-cut regular model. The set $G_s$ is called the
\emph{$s$-cut phase} of the family of Hermitian models and the decomposition~\eref{cou} is called the
\emph{phase diagram} of the family of Hermitian models.

\begin{figure}
  \begin{center}
   \includegraphics[width=10cm]{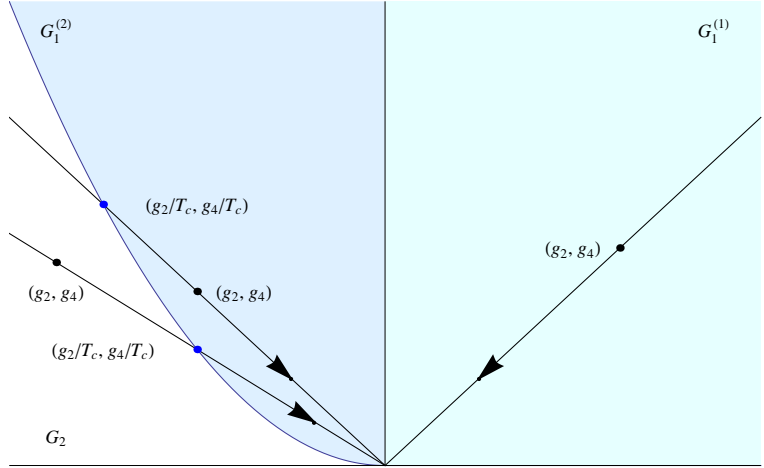}
   \end{center}
   \caption{Phase diagram of the quartic model $ V(\lambda) = g_2\lambda + g_4\lambda^2$ in the
   region $g_4>0$.\label{fig:sigma}}
\end{figure}

To illustrate these ideas we show in figure~\ref{fig:sigma} the phase diagram of the quartic model~\cite{BL08,BL99,BL03}
\begin{equation}
  \label{bip}
  V(\lambda) = g_2\lambda + g_4\lambda^2
\end{equation}
in the region
\begin{equation}
  G = \{\mathbf{g}=(g_2,g_4)\in \mathbf{R}^2: g_4>0 \}.
\end{equation}
The one-cut phase $G_1$ can be written as
\begin{equation}
  \label{ph1}
  G_1=G_1^{(1)}\cup G_1^{(2)},
\end{equation}
where
\begin{eqnarray}
  G_1^{(1)} = \{ (g_2,g_4)\in \mathbf{R}^2:  g_2\geq 0, g_4>0\},\\
  G_1^{(2)} = \{ (g_2,g_4)\in \mathbf{R}^2:  g_2<0, g_4>0,  g_2>-2\sqrt{g_4} \},
\end{eqnarray}
while the two-cut phase $G_2$ is given by
\begin{equation}
  G_2=\{ (g_2,g_4)\in \mathbf{R}^2:  g_4>0, g_2 < -2 \sqrt{g_4} \}.
\end{equation}
The phase diagram features the critical curve
\begin{equation}
  g_2=-2 \sqrt{g_4},
\end{equation}
which is the boundary between the two phases.

Figure~\ref{fig:sigma} also shows the oriented paths $\mathbf{g}/T$ traced
in the phase diagram when the coupling constants are scaled from a given initial point $\mathbf{g}$
by the temperature $T>0$ as $T\to\infty$. From the figure it is clear that if $\mathbf{g}\in G_1^{(1)}$
then $\mathbf{g}/T\in G_1$ for all $T>0$. However, if $\mathbf{g}\in G_1^{(2)}\cup G_2$  then $\mathbf{g}/T$
crosses the critical curve at
\begin{equation}
   \label{tcr}
   T_\mathrm{c} = \frac{g_2^2}{4 g_4}.
\end{equation}
\section{The discrete string equation}
Given $N>0$ let us denote by $v_n$ the orthogonal polynomials~\eref{mon}
\begin{equation}
   v_n(x) = P_{n,N}(x),\quad n\geq 0
\end{equation}
and let us denote by $L$ the Lax operator of the underlying Toda hierarchy~\cite{GE91}
\begin{equation}
  \label{lax}
  L v_n = v_{n+1} + r_{n,N} v_{n-1},\quad r_{0,N}=0.
\end{equation}

The main tool to study the asymptotics of the  recurrence coefficient $r_{n,N}$ is the \emph{discrete string equation}~\cite{BL08}
\begin{equation}
  \label{str01}
  V_z(L)_{n,n-1} = \frac{n}{N},
\end{equation}
where
\begin{equation}
  V_z(L) = \sum_{k=1}^{p} 2k g_{2k} L^{2k-1}
\end{equation}
and the subindices $(n,n-1)$ denote the matrix element between the corresponding elements of the basis $v_n$. 
We note that~\eref{str01} can be rewritten as
\begin{equation}
   \label{stri2}
   \oint_{\gamma}\frac{\rmd \lambda}{2\pi\rmi} V_{\lambda}(\lambda) U_{n,N}(\lambda) = \frac{n}{N},
\end{equation}
where $\gamma$ is a large positively oriented circle $|\lambda|=R$, and $U_{n,N}(\lambda)$ is the  generating function
\begin{equation}
  \label{gen}
  U_{n,N}(\lambda) = 1 + 2\sum_{k\geq 1} (L^{2k-1})_{n,n-1} \lambda^{-k}.
\end{equation}

In the next sections we calculate the large $N$ expansions of the recurrence coefficient $r_{n,N}$ by solving appropriate
continuum limits of the string equation in the form~\eref{stri2} with a two-step procedure:
(i) we first calculate a large $N$ expansion for the generating function $U_{n,N}$ whose coefficients are functions of
the expansion coefficients for $r_{n,N}$, and (ii) we substitute this expansion for $U_{n,N}$ into the
continuum limit of the string equation~\eref{stri2}, perform the contour integration (i.e.~pick up the coefficient
of $\lambda^{-1}$ in the large $\lambda$ expansion of the integrand), and solve recursively for the coefficients
of the $r_{n,N}$ expansion.

The first step relies on two identities for $U_{n,N}(\lambda)$  which we derive in the following theorem.
\begin{theo}
The generating function $ U_{n,N}(\lambda)$ satisfies the linear equation
\begin{equation}
  \label{lin}
  \lambda (U_{n+1,N}-U_{n,N}) = r_{n+1,N} (U_{n+2,N}+U_{n+1,N}) - r_{n,N} (U_{n,N}+U_{n-1,N})
\end{equation}
and the quadratic equation
\begin{equation}
  \label{res}
  r_{n,N} (U_{n,N}+U_{n-1,N}) (U_{n,N}+U_{n+1,N}) = \lambda (U_{n,N}^2-1).
\end{equation}
\end{theo}
\emph{Proof.} For clarity in this proof we will drop the subindex $N$ in $U_{n,N}$ and denote the generating
function~\eref{gen} by $U_n$. Let us prove first the linear equation~\eref{lin}. In terms of $L$ the recurrence relation~\eref{rec}
can be written as $L v_n(x)=x v_n(x)$. Therefore $L^{2k-1} v_n(x)=x^{2k-1} v_n(x)$ and
\begin{equation}
  (L^{2k-1})_{n,n-1}
  =
  \frac{1}{h_{n-1,N}}
  \int_{-\infty}^{\infty} x^{2k-1} v_n(x) v_{n-1}(x) \rmd\mu
\end{equation}
where $\rmd\mu = \rme^{-N V(x)} \rmd x$. Hence,
\begin{equation}
  \label{expu}
  U_n = 1+\frac{2}{h_{n-1,N}}
             \int_{-\infty}^{\infty} \frac{x v_n(x) v_{n-1}(x)}{\lambda-x^{2}}\rmd\mu
\end{equation}
and
\begin{eqnarray}
  \label{expu1}
  \fl
  \lambda (U_{n+1}-U_n)
  = & \frac{2}{h_{n,N}} \int_{-\infty}^{\infty} x v_{n+1}(x) v_{n}(x) \rmd \mu
       -\frac{2}{h_{n-1,N}} \int_{-\infty}^{\infty} x v_{n}(x) v_{n-1}(x) \rmd \mu
     \nonumber\\
     & {}+\frac{2}{h_{n,N}} \int_{-\infty}^{\infty} \frac{x^3 v_{n+1}(x) v_{n}(x)}{\lambda-x^{2}}\rmd \mu
           -\frac{2}{h_{n-1,N}} \int_{-\infty}^{\infty} \frac{x^3 v_n(x) v_{n-1}(x)}{\lambda-x^{2}}\rmd \mu.
\end{eqnarray}
Using $x^j v_n(x) =L^j v_n(x)$ for $j=1,2$ we find that
\begin{equation}
  \label{eq:sm}
  \int_{-\infty}^{\infty} x v_{n+1}(x) v_{n}(x) \rmd \mu = h_{n+1,N},
\end{equation}
\begin{eqnarray}
  \label{eq:la}
  \fl
  \int_{-\infty}^{\infty} \frac{x^3 v_{n+1}(x) v_{n}(x)}{\lambda-x^{2}}\rmd \mu  =
  \nonumber\\
 \fl
 \quad
 \int_{-\infty}^{\infty}
  \frac{x v_{n+2} v_{n+1} +x (r_{n+1,N}+r_{n,N}) v_n v_{n+1} + x r_{n,N}r_{n-1,N} v_{n+1} v_{n-2}}{\lambda-x^{2}}\rmd\mu,
\end{eqnarray}
and
\begin{eqnarray}
  \label{eq:lab}
    \fl\int_{-\infty}^{\infty} \frac{x^3 v_{n}(x) v_{n-1}(x)}{\lambda-x^{2}}\rmd \mu  =
  \nonumber\\
  \fl
  \quad
 \int_{-\infty}^{\infty}
  \frac{x v_{n+1} v_{n} +x r_{n,N}v_n v_{n-1}+x r_{n,N}r_{n-1,N}v_{n-1} v_{n-2}+x r_{n-1,N} v_{n+1}v_{n-2} }{\lambda-x^{2}}\rmd\mu.
\end{eqnarray}
Substituting~\eref{eq:sm}--\eref{eq:lab} into~\eref{expu1} and taking into account~\eref{expu} we conclude
that the linear equation~\eref{lin} holds. It is now easy to prove the quadratic equation~\eref{res}.
Indeed, the linear equation~\eref{lin} implies
\begin{equation}
  \fl
  r_{n+1,N} (U_{n+2}+U_{n+1}) (U_{n+1}+U_{n})
  -
  r_{n,N} (U_{n+1}+U_{n}) (U_{n}+U_{n-1})
  =
  \lambda (U_{n+1}^2-U_{n}^2),
\end{equation}
and therefore the expression
\begin{equation}
 \lambda U_{n}^2-r_{n,N} (U_{n+1}+U_{n}) (U_{n}+U_{n-1})
\end{equation}
is independent of $n$. Since $r_{0,N}=0$ and $U_0=1$ the quadratic equation~\eref{res} follows.
\section{Regular one-cut models}
In this section we show how to calculate large $N$ asymptotic expansions in regular one-cut models by passing
to the continuum limit in the quadratic identity~\eref{res}.
\subsection{Large $N$ expansions for regular one-cut models}
Consider the continuum limit
\begin{equation}
  \label{limit1}
  n\rightarrow \infty,\quad N\rightarrow \infty,\quad \frac{n}{N}\rightarrow T,
\end{equation}
in the case where $\mathbf{g}/T\in G_1$. It has been rigorously proved~\cite{DE99,BL05,KU00} that under these assumptions
there exists an asymptotic power series for the recurrence coefficient $r_{n,N}$ of the form
\begin{equation}
  \label{eq:con1}
  r_{n,N} \sim r(\epsilon,T)
                 =   \sum_{k\geq 0} r_k(T) \epsilon^{2k},\quad \epsilon=\frac{1}{N},
\end{equation}
which is uniform with respect to $T$ in a neighborhood of $T=1$. Moreover,
\begin{equation}
  \label{eq:r0}
  r_0 = \frac{\alpha^2}{4}
\end{equation}
where $(-\alpha,\alpha)$ is the eigenvalue support for the model with coupling constants $\mathbf{g}/T$.
We write a similar series for generating function $U_{n,N}$
\begin{equation}
  \label{genc}
  U_{n,N}(\lambda) \sim U(\lambda,\epsilon; r)
                             =    \sum_{k\geq 0} U_k(\lambda;r_0,\ldots,r_k) \epsilon^{2k},
\end{equation}
in which our notation for the coefficients $U_k$ anticipates their dependence on the $r_j$ up to $j=k$.
Substituting~\eref{eq:con1}, \eref{genc}~and the corresponding shifted expansions
\begin{equation}
   r_{n+j,N}\sim r_{[j]}(\epsilon,T) =  r(\epsilon,T+j\epsilon),\quad j\in \mathbf{Z},
\end{equation}
\begin{equation}
  U_{n+j,N}(\lambda) \sim U_{[j]}(\lambda,\epsilon;r) = U(\lambda,\epsilon;r_{[j]}),\quad j\in \mathbf{Z},
\end{equation}
into~\eref{res}, the continuum limit of the quadratic identity is
\begin{equation}
  \label{resc}
  r (U+U_{[-1]}) (U+U_{[1]}) = \lambda (U^2-1).
\end{equation}
Identifying powers of $\epsilon$ in~\eref{resc} we find that the coefficients $U_k$ can be written as
\begin{equation}
  \label{eq:u0}
  U_0 = \sqrt{\frac{\lambda}{\lambda-4r_0}},
\end{equation}
\begin{equation}
  \label{str}
  U_k = U_0\sum_{j=1}^{3k} \frac{ U_{k,j}(r_0,\ldots,r_k) }{ (\lambda-4r_0)^j },\quad  k\geq 1,
\end{equation}
where $U_{k,j}$ is a polynomial of degree $j$ in $r_0,\ldots, r_k$ and their $T$ derivatives.
For example,
\begin{equation}
 U_{1,1} = 2 r_1, \quad
 U_{1,2} = 2 r_0  r_0'', \quad
 U_{1,3} = 10 r_0 (r_0')^2.
\end{equation}
It is also easy to see that in general
\begin{equation}
  U_{k,1} = 2 r_k
\end{equation}
and that the dependence of $U_k$ in $ r_k$ comes solely from $U_{k,1}$. Therefore
\begin{equation}
  \label{rn}
  U_k = U_0 \left( \frac{2r_k}{\lambda-4r_0} + \cdots \right)
\end{equation}
where the dots stand for terms in $r_0,\ldots,r_{k-1}$ and their $T$ derivatives.

Analogously, the continuum limit of the string equation~\eref{stri2} reads
\begin{equation}
  \label{stri22}
  \oint_{\gamma}\frac{\rmd \lambda}{2\pi\rmi} V_{\lambda}(\lambda) U(\lambda,\epsilon;r) = T
\end{equation}
or using the expansion~\eref{genc} for $U$,
\begin{equation}
  \label{stri3c}
  \oint_{\gamma}\frac{\rmd \lambda}{2\pi\rmi}
  V_{\lambda}(\lambda) U_k(\lambda;r_0,\ldots,r_k)=\delta_{k,0}T,\quad k\geq 0.
\end{equation}
The first of these equations (corresponding to $k=0$) can be written as
\begin{equation}
  \label{sin0}
  W(r_0)=T,
\end{equation}
where
\begin{equation}
  \label{H}
  W(r_0) = \oint_{\gamma}\frac{\rmd \lambda}{2\pi\rmi}
                  V_{\lambda}(\lambda) \sqrt{\frac{\lambda}{\lambda-4 r_0}}
               = \sum_{k=1}^p {2 k \choose k} k  g_{2k}  r_0^k.
\end{equation}
Equation~\eref{sin0} is an implicit algebraic equation for $r_0$ of hodograph type
(linear in the variables $\mathbf{g}$ and $T$) which is equivalent to the normalization
condition~\eref{q1} due to the identification~\eref{eq:r0}.

The  remaining equations~\eref{stri3c} can be written as
 \begin{equation}
   \label{ache}
   \sum_{j=1}^{3 k} c_j(r_0)U_{k,j}(r_0,\ldots,r_k)=0,\quad k\geq 1,
\end{equation}
where
\begin{equation}
  \label{hj}
  c_j(r_0) = \oint_{\gamma}\frac{\rmd \lambda}{2\pi\rmi}
                      \frac{\sqrt{\lambda} V_{\lambda}(\lambda)}{(\lambda-4 r_0)^{j+\frac{1}{2}}}
                  = \frac{W^{(j)}(r_0)}{2^j (2j-1)!!}
\end{equation}
and where $W^{(j)}(r_0) = \partial^{j} W(r_0)/\partial r_0^{j}$.

For an even potential in the one-cut case $h(z)$ is an even function of $z$ whose characterization~\eref{0.3}
in terms of $\lambda=z^2$ reads
\begin{equation}
  2 \frac{ \sqrt{\lambda} V_{\lambda}(\lambda) }{ \sqrt{\lambda-\alpha^2} }
  =
  h(\lambda) + \Or(\lambda^{-1}),\quad \lambda\rightarrow \infty,
\end{equation}
and consequently
\begin{equation}
  \label{cri1}
  W'(r_0)
  =
  2 \oint_{\gamma}\frac{\rmd \lambda}{2\pi\rmi}
     \frac{\sqrt{\lambda} V_{\lambda}(\lambda)}{(\lambda-4 r_0)^{3/2}}
  = h(\lambda)\Big|_{\lambda=\alpha^2=4r_0}.
\end{equation}
Thus, if $\mathbf{g}/T\in G_1$ then $W'(r_0)\neq 0$ and the implicit function theorem
shows that~\eref{sin0} determines $r_0$ as a locally smooth function of $T$.
Moreover, from the dependency of $U_k$ on $r_k$ shown in~\eref{rn} it follows that
equations~\eref{ache} are of the form $r_k W'(r_0) = \cdots$, where the dots stand for
a sum of terms  on $r_0,\ldots,r_{k-1}$ and their $T$ derivatives. Therefore the implicit
function theorem can also be applied and~\eref{ache} determine recursively all the coefficients $r_k$
of the large $N$ expansion~\eref{eq:con1} as locally smooth functions of $T$.
Furthermore,  if we differentiate the hodograph equation~\eref{sin0} with respect to $T$ we can write
the $T$ derivatives of $r_k$ as rational functions of the $W^{(j)}$ and hence as rational functions of
$r_0$ and $\mathbf{g}$. For example, from $r'_0 = 1/W'(r_0)$ and the $k=1$ equation~\eref{ache}  we find
\begin{equation}
  \label{ere1}
  r_1  = \left(\frac{W''(r_0)^2}{6 W'(r_0)^4}-\frac{W'''(r_0)}{12 W'(r_0)^3}\right) r_0.
\end{equation}
\subsection{The quartic potential in the regular one-cut case}
As our first application of these results consider the quartic potential~\eref{bip} in the regular one-cut region.
Equation~\eref{H} gives
\begin{equation}
  \label{H4}
  W(r_0) = 2g_2r_0 + 12g_4r_0^2.
\end{equation}
The leading term $r_0(T)$ of the asymptotic expansion~\eref{eq:con1} is the positive root of the hodograph
equation $W(r_0) = T$, namely
\begin{equation}
  \label{eq:r0q}
  r_0(T) = \frac{\sqrt{g_2^2+12 T g_4}-g_2}{12 g_4}.
\end{equation}
To calculate the next term $r_1(T)$ we first use~\eref{ere1} and~\eref{H4} to obtain $r_1$ as a function of $r_0$,
and then we substitute the former explicit expression of $r_0(T)$ to get
\begin{equation}
   r_1(T) = \frac{g_4 (\sqrt{g_2^2+12 T g_4}-g_2)}{2(g_2^2+12 T g_4)^2}.
\end{equation}
Higher coefficients $r_k(T)$ of the asymptotic expansion~\eref{eq:con1} can be easily calculated iterating
this procedure. We postpone a graphical illustration of $r_0(T)$ until we discuss the two-cut region of the
same model in section~6.
\subsection{The Br\'ezin-Marinari-Parisi potential in the regular one-cut case}
\begin{figure}
  \begin{center}
   \includegraphics[width=10cm]{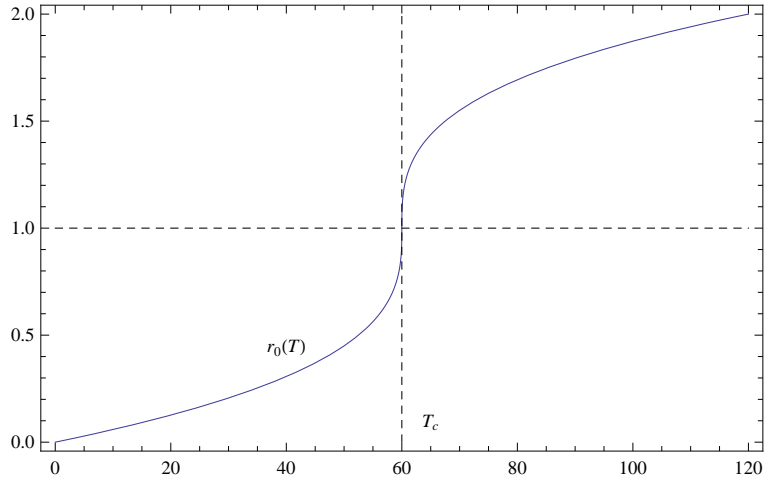}
   \end{center}
   \caption{Leading coefficient $r_0(T)$ as a function of $T$ for the Br\'ezin-Marinari-Parisi potential
                   $V(\lambda)= 90 \lambda-15 \lambda^2+\lambda^3$ with a critical point at $T_{\mathrm{c}}=60$.\label{fig:bmp}}
\end{figure}

As our second example we consider the sixtic potential
\begin{equation}
  \label{bmpp}
  V(\lambda)= 90 \lambda-15 \lambda^2+\lambda^3,
\end{equation}
introduced in~\cite{BR90} by Br\'ezin, Marinari and Parisi to generate a non-perturbative ambiguity-free
solution of a string model. The matrix model corresponding to $V(\lambda)/T$ is regular except at 
$T_{\mathrm{c}}=60$, but this critical temperature does not mark the boundary between a one-cut region and a
two-cut region. The function $W$ takes the form
\begin{equation}
  W(r_0) = 180 r_0 - 180 r_0^2 + 60 r_0^3,
\end{equation}
and the unique positive root of the hodograph equation $W(r_0)=T$ is given by
\begin{equation}
  \label{eq:r06}
  r_0(T) = 1+ \sqrt[3]{T/60-1}.
\end{equation}
Proceeding as in the previous example, equations~\eref{ere1} and~\eref{eq:r06} yield
\begin{equation}
 r_1(T) =  \frac{1+ \sqrt[3]{T/60-1}}{64800 (T/60-1)^2}.
\end{equation}
In figure~\ref{fig:bmp} we plot $r_0(T)$ as a function of $T$ in a neighborhood of the critical $T_{\mathrm{c}}=60$.
Note how the vertical tangent to the graph marks a critical point which is qualitatively different from the
graph of the quartic potential that we will discuss in section~6.
\section{Critical one-cut models and the Painlev\'e~I hierarchy }
In this section we first give a precise definition of singular solution of order $m$ in a one-cut model,
then we show how the algorithm to calculate the asymptotic expansion of the recurrence coefficient
developed in the previous section has to be modified by means of a suitable double scaling,
and finally we give a general proof of the relation between a one-cut singular model of order $m$
and the $(m-1)$-th Painlev\'e~I equation.
\subsection{Critical one-cut models and double scaling limit}
\begin{defi}
A solution  $(r_0,T)=(r_\mathrm{c},T_\mathrm{c})$ of the hodograph equation $ W(r_0) = T$ is called a singular
solution of order $m\geq 2$ if
\begin{equation}
  \label{mcr}
    W(r_\mathrm{c}) = T_\mathrm{c},
    \quad
    W'(r_\mathrm{c}) = \cdots = W^{(m-1)}(r_\mathrm{c}) = 0,
    \quad W^{(m)}(r_\mathrm{c}) \neq 0.
\end{equation}
\end{defi}

The identity
\begin{equation}
  \fl
  W^{(k)}(r_0)  = (2k-1)!! 2^k \oint_{\gamma}\frac{\rmd \lambda}{2\pi\rmi}
                                                            \frac{ \sqrt{\lambda} V_{\lambda}}{(\lambda-4 r_0)^{k+\frac{1}{2}}}
                          = \frac{(2k-1)!! 2^{k-1}}{(k-1)!} \frac{\partial^{k-1}}{\partial \lambda^{k-1}} h(\lambda)\Big|_{\lambda=4r_0}
\end{equation}
shows that singular solutions of order $m$ of the hodograph equation correspond to critical matrix models for
which the function $h(z)$ has zeros of order $m-1$  at the endpoints
of the  eigenvalue support $J= (-\alpha,\alpha)$. We note also that the eigenvalue
density $\rho(x)$ verifies
\begin{equation}
  \label{prh1}
  \rho(x) \sim (x\mp\alpha)^{m-\frac{1}{2}},\quad x\rightarrow \pm \alpha.
\end{equation}
These critical matrix models correspond to conformal $(2m-1,2)$ minimal models~\cite{DI95,BER09}.

Let us thus consider a model that (locally) for $T<T_\mathrm{c}$ is in the regular one-cut case and that for the critical
temperature $T=T_\mathrm{c}$ the corresponding hodograph equation~\eref{sin0} has a singular solution $r_\mathrm{c}$
of order $m\geq 2$. Since $W'(r_\mathrm{c})=0$ we cannot invoke the implicit function theorem to solve~\eref{sin0} near
$T=T_\mathrm{c}$ with $r_0(T_\mathrm{c})=r_\mathrm{c}$. In fact, $r_0(T)$ can be expanded in powers of
$(T-T_\mathrm{c})^{1/m}$ and therefore the system~\eref{stri3c} does not  determine  the coefficients $r_k$
of the expansion~\eref{eq:con1} as locally smooth functions of $T$ at $T=T_c$.
 We regularize this critical behavior introducing a scaling variable $x$ and a scaled expansion
parameter $\bar{\epsilon}$ defined by
\begin{equation}
  \label{eq:eps1c}
  \bar{\epsilon} = \epsilon^{1/(2m+1)} = N^{-1/(2m+1)},
\end{equation}
\begin{equation}
  \label{sca}
  T = T_\mathrm{c} + \bar{\epsilon}^{2m} x.
\end{equation}
In terms of these scaled variables the string equation~\eref{stri22} reads
\begin{equation}
  \label{stri22bc}
  \oint_{\gamma}\frac{\rmd \lambda}{2\rmi\pi} V_{\lambda}(\lambda) U(\lambda,\bar{\epsilon};r)
  = T_\mathrm{c} + \bar{\epsilon}^{2m} x,
\end{equation}
and we look for an asymptotic expansion of the form
 \begin{equation}
  \label{con1}
  r(\bar{\epsilon},x) = r_\mathrm{c} + \sum_{k\geq 1} \rt_k(x)  \bar{\epsilon}^{2k}.
\end{equation}
The shifts $T\rightarrow T\pm \epsilon$ correspond to $x\rightarrow x\pm \bar{\epsilon}$ and therefore
$U(\lambda,\bar{\epsilon};r)$ is determined by the quadratic equation
\begin{equation}
  \label{resca}
  r (U+U_{[\overline{-1}]}) (U+U_{[\bar{1}]}) = \lambda (U^2-1),
\end{equation}
where we are denoting $f_{[\bar{k}]}(x)=f(x+k \bar{\epsilon})$.

The rest of the calculation is straightforward and can be carried out in complete analogy with the
previous section. The corresponding expansion of $U$ is
\begin{equation}
  \label{stra}
 U(\lambda,\epsilon;\rt) = \sum_{k\geq 0} U^{[k]}(\lambda,{\rt}_1,\ldots,{\rt}_k)  \bar{\epsilon}^{2k}.
 \end{equation}
Substituting~\eref{con1} and~\eref{stra} in~\eref{resca} and equating powers of $\bar{\epsilon}$ we find that
\begin{equation}
  \label{uoc}
  U^{[0]} = \sqrt{\frac{\lambda}{\lambda-4 r_\mathrm{c}}},
\end{equation}
\begin{equation}
  \label{strab}
  U^{[k]} = U^{[0]}
  \sum_{j=1}^ k \frac{U^{[k,j]}({\rt}_{1},\ldots,{\rt}_{k-j+1})}{(\lambda-4 r_\mathrm{c})^j},\quad k\geq 1
\end{equation}
and  that the coefficients  $U^{[k,j]}$ are polynomials in ${\rt}_1,\ldots,{\rt}_{k-j+1}$
and their $x$ derivatives, which can be determined recursively~\cite{MA07}. In particular
\begin{equation}
 U^{[k,1]}= 2 \rt_k.
\end{equation}
The first few $U^{[k,j]}$ are
\begin{eqnarray}
   U^{[2,2]} = 6 \rt_1^2 + 2 r_\mathrm{c} \rt_1'',\\
  U^{[3,2]} = 12 \rt_1 \rt_2 +  2\rt_1 \rt_1'' + 2r_\mathrm{c} \rt_2''+ \frac{1}{6} r_\mathrm{c} \rt_1^{(4)},\\
  U^{[3,3]} = 20 \rt_1^3 + 10 r_\mathrm{c} (\rt_1')^2 + 20 r_\mathrm{c} \rt_1 \rt_1'' + 2 r_\mathrm{c}^2 \rt_1^{(4)}.
\end{eqnarray}
 In the following theorem we prove that the diagonal coefficients $U^{[k,k]}(\rt_1)$ are the well-known Gel'fand-Dikii differential
polynomials of the KdV theory  and  that
the coefficient $\rt_1(x)$  is a solution
of a member of the Painlev\'e~I equation.
\begin{theo}
  The coefficient $u=\rt_1$ for a singular one-cut model of  order $m$ satisfies the $(m-1)$-th Painlev\'e I equation
  ($P_I^{m-1}$ equation)
  \begin{equation}\label{epain}
  c_m(r_\mathrm{c})  R_m(u) = x,
  \end{equation}
where $R_m(u)$ is the differential polynomial in $u$ determined recursively  by
\begin{equation}
  \label{pain}
  \partial_x R_{m+1} = (r_\mathrm{c} \partial_x^3 + 4 u \partial_x + 2 u_x) R_m,\quad R_0=1,
\end{equation}
and
\begin{equation}
 c_m(r_\mathrm{c}) = \frac{W^{(m)}(r_\mathrm{c})}{2^m (2m-1)!!}.
\end{equation}
\end{theo}
\noindent
\emph{Proof}. The linear equation
\begin{equation}
  \label{linsa}
  r_{[\bar{1}]} (U_{[\bar{2}]}+U_{[\bar{1}]}) - r (U+U_{[\overline{-1}]}) = \lambda (U_{[\bar{1}]}-U)
\end{equation}
follows immediately from the quadratic equation~\eref{resca}, and substituting the expansion for $U$ given
by~\eref{stra} and~\eref{strab} into~\eref{linsa} we find that the diagonal coefficients $U^{[k,k]}$ satisfy
the recursion relation of the Gel'fand-Dikii differential polynomials,
\begin{equation}
  \partial_x U^{[k+1,k+1]}
  =
  (r_\mathrm{c} \partial_x^3 + 4 \rt_1 \partial_x + 2 (\rt_1)_x ) U^{[k,k]},\quad U^{[0,0]}=1.
\end{equation}
If we now substitute the expansion~\eref{stra} for $U$ into the scaled string equation~\eref{stri22bc},
equate powers of $\bar{\epsilon}$, and take into account~\eref{hj} as well as the definition of singular solution~\eref{mcr},
we obtain
\begin{equation}
  \label{straab}
  \sum_{j=m}^k c_j(r_\mathrm{c}) U^{[k,j]}(\rt_1,\ldots,\rt_{k-j+1}) = \delta_{k,m} x,
  \quad
  k\geq m,
\end{equation}
so that
\begin{eqnarray}
  \label{p}
  c_m(r_\mathrm{c}) U^{[m,m]}(\rt_1) = x,\\
  \label{mp}
  \sum_{j=m}^{m+k}c_j(r_\mathrm{c}) U^{[m+k,j]}(\rt_1,\ldots,\rt_{m+k-j+1}) = 0,\quad k\geq 1.
\end{eqnarray}
This system provides for each coefficient $\rt_k(x)\,(k\geq 1)$
an ordinary differential equation involving the previous coefficients $ \rt_j(1\leq j <k)$. 
In particular~\eref{p} takes the form~\eref{epain} as stated. The set of ordinary differential
equations~\eref{epain}  constitutes the Painlev\'e I hierarchy, whose $m=2$ member
is the familiar Painlev\'e~I equation
\begin{equation}
  \label{paine}
  c_2(r_\mathrm{c}) (2 r_\mathrm{c} u_{xx}+6 u^2)=x.
\end{equation}

Note that a more precise notation for $R_m(u)$ and $c_m(r_\mathrm{c})$ in theorem~2 would be $R_m(r_\mathrm{c},u)$ and
$c_m(r_\mathrm{c},\mathbf{g})$, which emphasizes the explicit dependence on the critical value $r_\mathrm{c}$
and the coupling constants $\mathbf{g}$.
\subsection{The critical Br\'ezin-Marinari-Parisi model }
We can apply immediately Theorem~2 to the Br\'ezin-Marinari-Parisi potential.
The singular solution $(r_\mathrm{c}=1,T_\mathrm{c} =60)$ of the corresponding hodograph equation is of third order, since
\begin{equation}
  W(r_c)  =  T_c,
  \quad
  W'(r_c) = 0,
  \quad W''(r_c) = 0,
  \quad W'''(r_c) = 360.
\end{equation}
Therefore $u={\rt}_1$ is as solution of the $P_I^2$ equation
\begin{equation}
  \label{pbmp}
  u_{xxxx} + 10 u u_{xx} + 5 u_x^2 + 10 u^3 = \frac{1}{6}x.
\end{equation}
\section{Regular two-cut models}
In this section we study regular two-cut models (i.e.~$\mathbf{g}/T\in G_2$ for a given $T>0$) following the same pattern
used in our previous study of regular one-cut models. The main difference is that now, due to~\eref{asy1},
we need two distinct asymptotic expansions: one expansion for the odd recurrence coefficients $r_{2n+1,N}$ and a second
expansion for the even recurrence coefficients $r_{2n,N}$. This fact leads us to introduce two different generating
functions $\bV$ and $\bW$ for  $U_{n,N}$ with $n$ odd and even, respectively. It also requires the splitting of the quadratic
equation~\eref{res} for $U_{n,N}$ into a system of two coupled equations.
\subsection{Large $N$ expansions for regular two-cut models}
In view of~\eref{asy1}  we formulate the asymptotics of $r_{n,N}$ in the limit
\begin{equation}
  \label{eq:limit2}
  n\rightarrow \infty,\quad N\rightarrow \infty,\quad \frac{2n}{N}\rightarrow T
\end{equation}
as
\begin{equation}
  \label{con2}
  r_{2n+1,N}\sim a_{[1]}(\epsilon,T),\quad
  r_{2n,N}     \sim b(\epsilon,T),
\end{equation}
where we use again the bracket notation for shifts and we assume
that $a(\epsilon,T)$ and $b(\epsilon,T)$ are asymptotic power series
 \begin{equation}\label{a,b}
   a(\epsilon,T) = \sum_{k\geq 0} a_k(T)\epsilon^{2k},\quad
   b(\epsilon,T) = \sum_{k\geq 0} b_k(T)\epsilon^{2k},\quad
   \epsilon = \frac{1}{N}.
 \end{equation}
We notice that contrarily to the one-cut case and with the exception of the
quartic model~\cite{BL08,BL99,BL03}, to our knowledge there is no rigorous proof that power series
with smooth coefficients are truly asymptotic to the recurrence coefficients.

To calculate the coefficients $a_k$ and $b_k$ we perform the continuum limit~\eref{eq:limit2}
in the string and quadratic equations~\eref{stri2} and~\eref{res}, and introduce two generating functions
\begin{eqnarray}
  \label{R,S}
  \eqalign{
    \mathrm{V}(\lambda,\epsilon;a,b) = \sum_{k\geq 0} \bV_k(\lambda;a_0,b_0,\ldots,a_k,b_k) \epsilon^{2k},\cr
    \mathrm{W}(\lambda,\epsilon;a,b) = \sum_{k\geq 0} \bW_k(\lambda;a_0,b_0,\ldots,a_k,b_k) \epsilon^{2k},}
\end{eqnarray}
such that
\begin{equation}
  \label{con22}
  U_{2 n+1,N}(\lambda)\sim \bV_{[1]}(\lambda,\epsilon;a,b),\quad
  U_{2 n,N}(\lambda)\sim  \bW(\lambda,\epsilon;a,b),
\end{equation}
and more generally
\begin{equation}
  \label{con22cc}
  U_{2(n+j)+1,N}(\lambda)\sim \bV_{[2j+1]}(\lambda,\epsilon;a,b),\quad
  U_{2(n+j),N}(\lambda)\sim  \bW_{[2j]}(\lambda,\epsilon;a,b).
\end{equation}
Then~\eref{res} is equivalent to the system
\begin{equation}
  \label{resc2a}
  \eqalign{
    a  (\bV+ \bW_{[-1]}) (\bV+ \bW_{[1]}) = \lambda  (\bV^2-1),\cr
    b  ( \bW+\bV_{[-1]}) ( \bW+\bV_{[1]}) = \lambda  (\bW^2-1).}
\end{equation}
Identifying the coefficients of $\epsilon^0$ in these equations we get
\begin{equation}
  \label{r0}
  \bV_0 = \frac{a_0-b_0+\lambda}{w},\quad
  \bW_0 = \frac{b_0-a_0+\lambda}{w},
\end{equation}
where
\begin{equation}
 w = \sqrt{ \lambda^2-2\lambda(a_0+b_0)+(b_0-a_0)^2}
    = \sqrt{(\lambda-\alpha^2)(\lambda-\beta^2)}.
\end{equation}
The last equality follows from~\eref{asy1}, which allows us to express $a_0$ and $b_0$
in terms of the endpoints of the eigenvalue support $J=(-\beta,-\alpha)\cup (\alpha,\beta)$
of the model with coupling constants $\mathbf{g}/T$:
\begin{equation}
  \label{aes}
  a_0=\frac{1}{4} (\alpha+\beta)^2,\quad b_0=\frac{1}{4}(\alpha-\beta)^2.
\end{equation}

Recursive identification of the coefficient of $\epsilon^k$ for $k>0$ in~\eref{resc2a} leads
to a system of two linear equations in $\bV_k$ and $\bW_k$ of the form
\begin{equation}
  \label{resc2ac0}
  \eqalign{
                  (\lambda \bV_0-2{w^{-1}}\lambda a_0)\bV_{k} - 2{w^{-1}}\lambda a_0\bW_{k}=\cdots\cr
                  -2{w^{-1}}\lambda b_0 \bV_{k} + (\lambda \bW_0-2{w^{-1}}\lambda b_0)\bW_{k}=\cdots}
\end{equation}
where the right-hand sides are functions of $a_0,b_0,\ldots,a_k,b_k$ and their $T$ derivatives.
Since the determinant of the coefficients of~\eref{resc2ac0} equals $\lambda^2$ we can solve
uniquely~\eref{resc2ac0} for $\bV_k$ and $ \bW_k$ as functions of $a_0,b_0,\ldots,a_k,b_k$ and
their $T$ derivatives. Moreover, if $\bV(\lambda,\epsilon;a,b)$ and $\bW(\lambda,\epsilon;a,b)$
satisfy~\eref{resc2a}, so do the functions
\begin{equation}
  \widetilde{\bV}(\lambda,\epsilon;a,b) = \bW(\lambda,\epsilon;b,a),
  \quad
  \widetilde{\bW}(\lambda,\epsilon,a,b) = \bV(\lambda,\epsilon;b,a).
\end{equation}
From these equations and taking into account that as $\lambda\rightarrow\infty$
\begin{equation}
  \label{asy}
  \bV(\lambda,\epsilon;a,b) \sim 1+ \Or(\lambda^{-1}),\quad
  \bW(\lambda,\epsilon;a,b)\sim 1+ \Or(\lambda^{-1}),
\end{equation}
it follows that $\bV$ and $ \bW$ are related by
\begin{equation}\label{sim}
  \bW(\lambda,\epsilon;a,b)= \bV(\lambda,\epsilon;b,a).
\end{equation}

Using induction in the quadratic system~\eref{resc2a} we find that  the coefficients $\bV_k$ can be written in the form
\begin{equation}\label{rkj}
  \bV_k = \sum_{j=0}^{3k} \frac{ R_{k,j} +\lambda S_{k,j}}{ w^{2j+1} },\quad  k\geq 0,
\end{equation}
where $R_{k,j}=R_{k,j}(a_0,b_0,\ldots,a_k,b_k)$ and $S_{k,j}=S_{k,j}(a_0,b_0,\ldots,a_k,b_k)$
are polynomials in $a_0,b_0,\ldots,a_k,b_k$ and their $T$ derivatives.
 It is immediate to prove
 that the dependence of $\bV_k$
on $(a_k,b_k)$ is given by
\begin{equation}
  \label{rn22}
    \bV_k = \frac{2a_k}{w} +\frac{(2\lambda (a_0+b_0)-2(b_0-a_0)^2) a_k+4\lambda a_0 b_k}{w^{3}} +
             \cdots,\quad k\geq 1,
\end{equation}
where the dots stand for terms dependent on $a_0,b_0,\ldots,a_{k-1},b_{k-1}$  and their $T$ derivatives.

The next step is to calculate the $a_k$ and $b_k$ as functions of $T$. Substituting the expressions for
$\bV$ and $\bW$ into the continuum limit of the string equation~\eref{stri2} it splits into the
system of equations
\begin{equation}
  \label{stri2c2r}
  \eqalign{\oint_{\gamma}\frac{\rmd \lambda}{2\pi\rmi} V_{\lambda}(\lambda) \bV(\lambda,\epsilon;a,b)=T,\cr
                \oint_{\gamma}\frac{\rmd \lambda}{2\pi\rmi} V_{\lambda}(\lambda) \bV(\lambda,\epsilon;b,a)=T,}
\end{equation}
or, equivalently,
\begin{equation}
  \label{stri3c2r}
  \eqalign{\oint_{\gamma}\frac{\rmd \lambda}{2\pi\rmi}
                V_{\lambda}(\lambda) \bV_k(\lambda;a_0,b_0,\ldots,a_k,b_k) = \delta_{k,0}T,\cr
                \oint_{\gamma}\frac{\rmd \lambda}{2\pi\rmi}
                V_{\lambda}(\lambda) \bV_k(\lambda;b_0,a_0,\ldots,b_k,a_k) = \delta_{k,0}T.}
\end{equation}
For $k=0$ we have a system of two hodograph equations for $a_0(T)$ and $b_0(T)$,
\begin{equation}
  \label{imp21}
  W(a_0,b_0) = T,\quad W(b_0,a_0)=T,
\end{equation}
where
\begin{equation}\label{w2cu}
  W(a_0,b_0) = \oint_{\gamma}\frac{\rmd \lambda}{2\pi\rmi}
                                                            V_{\lambda}(\lambda) \frac{\lambda+a_0-b_0}{w}.
\end{equation}
Note that in view of~\eref{aes} the equations~\eref{imp21} are equivalent to the equations~\eref{q2}
which determine the two-cut support for a model with coupling constants $\mathbf{g}/T$.

Finally, we say that $(a_0,b_0)$ is a \emph{regular} solution of~\eref{imp21} if the Jacobian determinant
of~\eref{imp21} with respect to $(a_0,b_0)$ does not vanish. Otherwise $(a_0,b_0)$ is called a \emph{singular} solution.
Equation~\eref{rn22} shows that if $(a_0,b_0)$  is a  regular solution of~\eref{imp21}, then the system~\eref{stri3c2r}
determines recursively all the coefficients $a_k$ and $b_k$ as locally smooth functions of $T$.
\subsection{The quartic model in the regular two-cut case}
As we have discussed at the end of section~2, for $g_2<0$ and $T < T_c = g_2^2/(4g_4)$ the quartic model~\eref{bip}
is in the regular two-cut case, and we can calculate expansions of the form~\eref{a,b}. The first coefficient $\bV_0$
can be read off directly from~\eref{r0}. Thus, the string equations~\eref{stri3c2r} for $k=0$ are
\begin{equation}
  \eqalign{
  2 g_2 a_0+4 g_4 (a_0^2+2 b_0a_0) = T,\cr
  2 g_2 b_0+4 g_4 (b_0^2+2 a_0b_0) = T,}
\end{equation}
and we find immediately the leading coefficients
\begin{eqnarray}
  \label{eq:a0q}
  a_0(T) = \frac{\sqrt{g_2^2-4 T g_4}-g_2}{4 g_4},\\
  \label{eq:b0q}
  b_0(T) = \frac{-\sqrt{g_2^2-4 T g_4}-g_2}{4 g_4}.
\end{eqnarray}

\begin{figure}
  \begin{center}
   \includegraphics[width=10cm]{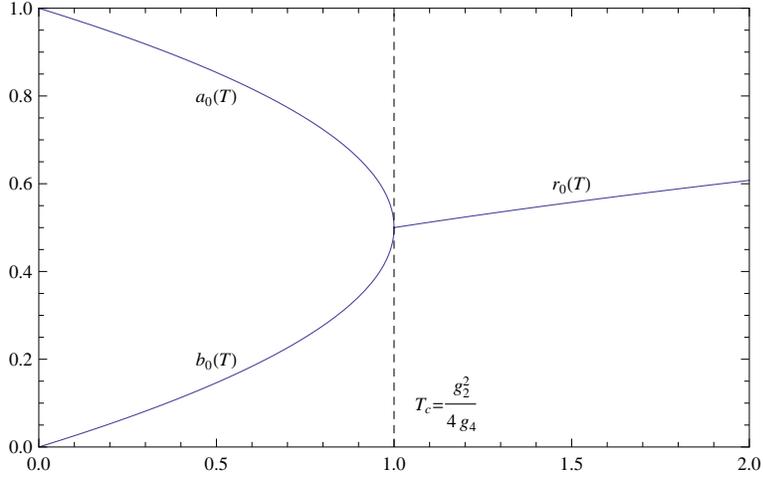}
   \end{center}
   \caption{Leading terms $r_0$, $a_0$ and $b_0$ as a function of $T$ for the quartic model
   $ V(\lambda) = g_2\lambda + g_4\lambda^2$ with $g_2=-2$ and $g_4=1$.\label{fig:r0a0b0}}
\end{figure}

Then we solve~\eref{resc2a} for the next term $\bV_1$, of which in the notation of equation~\eref{rkj}
we need the following coefficients:
\begin{equation}
  R_{1,0} = 2a_1,\quad
  S_{1,0} = 0,\quad
  S_{1,1} = 2(a_0+b_0)a_1+4a_0b_1+2a_0b_0''.
\end{equation}
With these expressions we find that the string equations~\eref{stri3c2r} for $k=1$ are
\begin{equation}
\eqalign{
  2 g_2a_1+2 g_4 (4 a_0a_1+4 b_0a_1+4 a_0b_1+2a_0b_0'') = 0,\cr
  2 g_2b_1+2 g_4 (4 b_0b_1+4 a_0b_1+4 b_0a_1+2b_0a_0'') = 0.}
\end{equation}
Hence, the second coefficients are
\begin{eqnarray}
  a_1(T) = -\frac{g_4 (g_2^2+4 T g_4-g_2\sqrt{g_2^2-4 T g_4})}{
                       2 (g_2^2-4 T g_4)^{5/2}},\\
  b_1(T) = \frac{g_4(g_2^2+4 T g_4+g_2\sqrt{g_2^2-4 T g_4})}{
                      2 (g_2^2-4 T g_4)^{5/2}}.
\end{eqnarray}
In figure~\ref{fig:r0a0b0} we illustrate these asymptotic behaviors for a quartic model with $g_2=-2$ and $g_4=1$,
so that the critical temperature is $T_\mathrm{c}=1$. For $0<T<T_\mathrm{c}$ we plot the leading coefficients
$a_0(T)$ and $b_0(T)$ of the asymptotic expansions for the odd and even terms of the recursion coefficient in
the two-cut region given by equations~\eref{eq:a0q} and~\eref{eq:b0q} respectively, while for $T>T_\mathrm{c}$ we plot
the leading coefficient $r_0(T)$ of the asymptotic expansion in the one-cut region given by equation~\eref{eq:r0q}.
As the temperature increases through $T_\mathrm{c}$ the model features a merging of two cuts~\cite{BL99}.
In fact this is the simplest example of the general processes that we classify in the next section.
\section{Critical two-cut models with merging of two cuts}
In order to classify  critical models in the two-cut case it is useful to consider the function
\begin{equation}\label{F}
F(a_0,b_0)=\oint_{\gamma}\frac{\rmd \lambda}{2\pi\rmi}
                                            V_{\lambda}(\lambda) w(\lambda,a_0,b_0) + T (a_0+b_0) ,
\end{equation}
where
\begin{equation}
 w(\lambda,a_0,b_0)^2= \lambda^2-2\lambda(a_0+b_0)+(b_0-a_0)^2,
\end{equation}
because the solutions of the system
\begin{equation}
  \label{crF}
  \frac{\partial F}{\partial a_0} = 0,
  \quad
  \frac{\partial F}{\partial b_0} =0,
\end{equation}
are precisely the solutions of the hodograph system~\eref{imp21}. Moreover, it is convenient to change variables
from $(a_0,b_0)$ to a new pair $(\sigma,\tau)$ defined by
\begin{equation}
  \label{tet}
  \sigma = \alpha^2 = a_0 + b_0 - 2 \sqrt{a_0 b_0},
  \quad
  \tau= \beta^2  = a_0 + b_0 + 2 \sqrt{a_0 b_0},
\end{equation}
where $\alpha$ and $\beta$ determine the endpoints of the eigenvalue support
$J$ (cf.~\eref{aes}). In these new variables the function~\eref{F} is given by
\begin{equation}
  \label{Fn}
  F(\sigma,\tau) = \oint_{\gamma}\frac{\rmd \lambda}{2\pi\rmi}
                                             V_{\lambda}(\lambda)\sqrt{(\lambda-\sigma)(\lambda-\tau)}
                                             + \frac{T}{2}(\sigma+\tau),
\end{equation}
and satisfies the Euler-Poisson-Darboux equation~\cite{DA15}
\begin{equation}
  \label{epd}
  2 (\tau-\sigma) \frac{\partial^2 F}{\partial \sigma \partial \tau}
  =
  \frac{\partial F}{\partial \sigma}-\frac{\partial F}{\partial \tau}.
\end{equation}
Hence it is clearly advantageous to analyze the solutions $(a_0,b_0)$ of the hodograph
system~\eref{crF} in terms of the solutions $(\sigma,\tau)$ of the transformed system
\begin{equation}
  \label{crF2}
  \frac{\partial F}{\partial \sigma} = 0,
  \quad
  \frac{\partial F}{\partial \tau} =0,
\end{equation}
because for any solution $(\sigma,\tau)$ of~\eref{crF2} with  $\sigma\neq\tau$
 the Euler-Poisson-Darboux equation permits to express any mixed derivative
$\partial^{i+j} F/\partial\sigma^i\partial\tau^j$ as a linear combination of unmixed derivatives
$\partial^n F/\partial\sigma^n$ and  $\partial^m F/\partial\tau^m$.
\subsection{Merging of two cuts}
We are interested in solutions of~\eref{crF2} with $\sigma= 0$ (i.e. $a_0= b_0$). Since
\begin{eqnarray}
  \frac{\partial F}{\partial a_0}
  = \left(1-\sqrt{\frac{b_0}{a_0}}\right) \frac{\partial F}{\partial \sigma} + 
     \left(1+\sqrt{\frac{b_0}{a_0}}\right) \frac{\partial F}{\partial \tau},\\
  \frac{\partial F}{\partial b_0}
  = \left(1-\sqrt{\frac{a_0}{b_0}}\right) \frac{\partial F}{\partial\sigma} + 
    \left(1+\sqrt{\frac{a_0}{b_0}}\right) \frac{\partial F}{\partial \tau},
\end{eqnarray}
we have that
\begin{equation}
  \left|\begin{array}{cc}
    \displaystyle \frac{\partial^2 F}{\partial a_0^2} & \displaystyle\frac{\partial^2 F}{\partial a_0\partial b_0}\\
    \\
    \displaystyle \frac{\partial^2 F}{\partial a_0\partial b_0} & \displaystyle\frac{\partial^2 F}{\partial b_0^2}
  \end{array}\right|
  =
  32 \frac{\partial^2F}{\partial\tau^2}(0,\tau) \frac{\partial F}{\partial \sigma}(0,\tau)=0.
\end{equation}
Therefore these solutions  of~\eref{crF2}  with $\sigma=0$ determine singular solutions of~\eref{crF}. Moreover,
the corresponding matrix models have an eigenvalue support with $\alpha=0$ so that they represent
a critical process of merging of two cuts. This motivates our next definition:
\begin{defi}
A solution   $(\sigma,\tau)$ of~\eref{crF2} with $\sigma=0$  and $\tau>0$ determines a singular solution
of the system~\eref{crF} with a merging of two cuts of order $m\geq 1$ if
\begin{equation}
  \label{m12}
  \eqalign{
  \frac{\partial^{k} F}{\partial \sigma^{k}}(0,\tau) =0 ,\quad  k=1,\ldots,m;\quad   \frac{\partial^{m+1}F}{\partial\sigma^{m+1}}(0,\tau)\neq 0,\cr
  \frac{\partial F}{\partial \tau}(0,\tau) =0,\quad   \frac{\partial^{2}F}{\partial\tau^2}(0,\tau)\neq 0.}
  \end{equation}
\end{defi}
For these solutions of~\eref{crF}  we have that
\begin{eqnarray}
  \frac{\partial^{k+1} F}{\partial \sigma^{k+1}}(0,\tau)
  &= -\frac{(2k-1)!!}{2^{k+1}}
        \oint_{\gamma}\frac{\rmd \lambda}{2\pi\rmi}
        \frac{(\lambda-\tau) V_{\lambda}}{\lambda^k  w}\nonumber\\
  &= -\frac{(2k-1)!!}{2^{k+1}(k-1)!}
        \left.
        \frac{\partial^{k-1}}{\partial \lambda^{k-1}}
        \left((\lambda-\tau)\widetilde{h}(\lambda)\right)\right|_{\lambda=0},\quad k\geq 1,
\end{eqnarray}
where $\widetilde{h}$ is the polynomial in $\lambda=z^2$ given by
\begin{equation}
  \widetilde{h}(\lambda) = \frac{h(z)}{2 z}.
\end{equation}
Thus $h(z)$ has a zero of order $2m-1$ at $z=0$ and the eigenvalue density $\rho(x)$ verifies
\begin{equation}
  \label{prh}
  \rho(x) \sim x^{2m},\quad x\rightarrow 0.
\end{equation}
These critical models are related to conformal $(2m,1)$ minimal models~\cite{BER09,MA10}.

Note that there is a difference between the definitions of order of a singular solution in the one-cut (definition~1) and
two-cut (definition~2) cases, which in turn entails a difference in the statements of our main results (theorems~2 and~3).
However, with these definitions we achieve a complete analogy in the form of both scaling expansion parameters
$\bar{\epsilon}$~\eref{eq:eps1c} and~\eref{eq:eps2c} as well as the corresponding string equations~\eref{stri22bc}
and~\eref{stri2c2rcc}.
\subsection{ Merging of two cuts in the  quartic model}
As an illustration of these ideas, consider the quartic model~\eref{bip}. The function~\eref{Fn} is
\begin{equation}
  \fl
  F(\sigma,\tau)
  =
  \frac{1}{8} ( - g_4 \sigma ^3 - g_2 \sigma ^2 + g_4\tau \sigma ^2 + 2g_2  \tau  \sigma
                      + g_4\tau ^2  \sigma - g_2\tau ^2 - g_4\tau ^3) + \frac{1}{2} T (\sigma +\tau ).
\end{equation}
It follows at once that at the critical temperature $T_\mathrm{c}=g_2^2/(4g_4)$ we have a solution $(0,\tau_\mathrm{c})$
of the system
\begin{equation}
  \frac{\partial F}{\partial\sigma}(0,\tau_\mathrm{c}) = 0,
  \quad
  \frac{\partial F}{\partial\tau}(0,\tau_\mathrm{c}) = 0,
\end{equation}given by $\tau_\mathrm{c}=-g_2/g_4$. Moreover, since
\begin{equation}
  \frac{\partial^2 F}{\partial\sigma^2}(0,\tau_\mathrm{c}) = -\frac{g_2}{2} \neq 0,
\end{equation}
this solution describes a merging of two cuts of order $m=1$ .
\subsection{Double-scaling limit and string equations}
Let us consider a matrix model $V(\lambda)/T$ with a critical point at $T=T_\mathrm{c}$ such that locally for
$T<T_\mathrm{c}$ the model is a regular two-cut model, and at $T=T_\mathrm{c}$ the model features a singular
two-cut merging of order $m$ as described in the previous subsection. Then
\begin{equation}
  \lim_{T\rightarrow T_\mathrm{c}-0} a_0(T) = \lim_{T\rightarrow T_\mathrm{c}-0}b_0(T) = r_\mathrm{c},
\end{equation}
so that the two cuts of the eigenvalue support merge to $J=(-\beta_\mathrm{c},\beta_\mathrm{c}) $ where
$\beta_\mathrm{c}^2=4r_\mathrm{c}$, and the hodograph system~\eref{imp21} reduces to
\begin{equation}
  \label{e0}
  \eqalign{\oint_{\gamma}\frac{\rmd \lambda}{2\pi\rmi}
                V_{\lambda}(\lambda) \frac{\lambda}{w_\mathrm{c}}=T_\mathrm{c},\cr
                \oint_{\gamma}\frac{\rmd \lambda}{2\pi\rmi}
                V_{\lambda}(\lambda) \frac{1}{w_\mathrm{c}}= 0,}
\end{equation}
where
\begin{equation}
  \label{efec}
  w_\mathrm{c} =\sqrt{ \lambda (\lambda-4 r_\mathrm{c})}.
\end{equation}
Note that the hodograph system~\eref{e0} in effect determines both $T_\mathrm{c}$ and $r_\mathrm{c}$
in terms of the coupling parameters $\mathbf{g}$. Moreover, the definition~\eref{m12} of singular solution with a
merging of order $m$ reads
 \begin{equation}
   \label{idd}
  \oint_{\gamma}\frac{\rmd \lambda}{2\pi\rmi}
  \frac{(\lambda-4 r_\mathrm{c})^{k+1} V_{\lambda}}{w_\mathrm{c}^{2k+1}}=0,\quad k=1,\ldots,m-1,
\end{equation}
\begin{equation}
  \label{idd2}
  \oint_{\gamma}\frac{\rmd \lambda}{2\pi\rmi}
  \frac{(\lambda-4 r_\mathrm{c})^{m+1} V_{\lambda}}{w_\mathrm{c}^{2m+1 }}\neq 0,
  \quad
  \oint_{\gamma}\frac{\rmd \lambda}{2\pi\rmi}
  \frac{\lambda^2 V_{\lambda}}{w_\mathrm{c}^{3}}\neq 0.
\end{equation}
Equations~\eref{idd} and~\eref{idd2} characterize the subset of the phase space $G$ representing these critical models.

As in our study of critical one-cut models, the implicit function theorem cannot be applied to solve~\eref{stri22} near
$T=T_\mathrm{c}$,  with $a_0(T_\mathrm{c})=b_0(T_\mathrm{c})=r_\mathrm{c}$. In analogy with~\eref{sca}
we regularize this critical behavior in the continuum limit
\begin{equation}
  \label{limit2}
  n\rightarrow \infty,\quad N\rightarrow \infty,\quad \frac{2n}{N}\rightarrow T
\end{equation}
using a scaling variable $x$ and a scaled expansion parameter $\bar{\epsilon}$ defined by
\begin{equation}
 \label{eq:eps2c}
  \bar{\epsilon} = \epsilon^{1/(2m+1)} = N^{-1/(2m+1)},
\end{equation}
\begin{equation}
  \label{scabis}
  T = T_\mathrm{c} + \bar{\epsilon}^{2m} x.
\end{equation}
The continuum limit
of the recurrence coefficient can be written as
\begin{equation}
    r_{2n+1,N}\sim \at_{[\overline{1}]}(\bar{\epsilon},x),\quad
  r_{2n,N}     \sim \bt(\bar{\epsilon},x),
\end{equation}
where we denote $f_{[\bar{k}]}(x)=f(x+k\bar{\epsilon})$  and $\at(\bar{\epsilon},x)$
and $\bt(\bar{\epsilon},x)$ denote asymptotic power series
 \begin{equation}
   \label{a,b,c}
   \at(\bar{\epsilon},x) = r_\mathrm{c}+\sum_{k\geq 1} \at_k(x)\bar{\epsilon}^{k},\quad
   \bt(\bar{\epsilon},x) = r_\mathrm{c}+\sum_{k\geq 1} \bt_k(x)\bar{\epsilon}^{k}.
 \end{equation}
We introduce again two generating functions
\begin{equation}
  \label{RcS}
  \eqalign{\bV(\lambda,\bar{\epsilon};\at,\bt) = \sum_{k\geq 0} \bV^{[k]}(\lambda;\at_1,\bt_1,\ldots,\at_k,\bt_k) \bar{\epsilon}^{k},\cr
                \bW(\lambda,\bar{\epsilon};\at,\bt) = \sum_{k\geq 0} \bW^{[k]}(\lambda;\at_1,\bt_1,\ldots,\at_k,\bt_k) \bar{\epsilon}^{k},}
\end{equation}
such that the asymptotic behaviors of the odd and even terms of the generating function $U_n$ are given respectively by
\begin{equation}
  \label{con22c}
  U_{2n+1}(\lambda)\sim \bV_{[\overline{1}]}(\lambda,\bar{\epsilon};\at,\bt) ,\quad
  U_{2n}(\lambda)\sim  \bW(\lambda,\bar{\epsilon};\at,\bt).
\end{equation}
Thus the continuum limit of the quadratic equation~\eref{res} for $U_n$ splits into the system
\begin{equation}
  \label{resc2ac}
  \eqalign{ \at (\bV+ \bW_{[\overline{-1}]}) (\bV+ \bW_{[\overline{1}]}) = \lambda (\bV^2-1),\cr
                 \bt (\bW+\bV_{[\overline{-1}]}) ( \bW+\bV_{[\overline{1}]}) = \lambda (\bW^2-1).}
\end{equation}
Identification of the coefficients  of $\bar{\epsilon}^0$ in~\eref{resc2ac} leads to
\begin{equation}
  \bV^{[0]} = \bW^{[0]}= \frac{\lambda}{w_\mathrm{c}},
\end{equation}
and recursive identification of the coefficients of $\bar{\epsilon}^k$ for $k>0$ to a system of two linear
equations for $\bV^{[k]}$ and $\bW^{[k]}$ of the form
\begin{equation}
  \label{resc2ac2}
  \eqalign{ (2\lambda -4 r_\mathrm{c}) \bV^{[k]} - 4 r_\mathrm{c} \bW^{[k]} = \cdots,\cr
                 -4 r_\mathrm{c} \bV^{[k]} +  (2\lambda -4 r_\mathrm{c}) \bW^{[k]} = \cdots,}
\end{equation}
where the right-hand sides are functions of $\at_1,\bt_1,\ldots,\at_k,\bt_k$ and their $x$ derivatives.
Once more, the determinant of the coefficients does not vanish and~\eref{resc2ac}
uniquely determine the  coefficients  $\bV^{[k]}$ and $\bW^{[k]}$  as polynomials in the functions
$\at_1,\bt_1,\ldots,\at_k,\bt_k$ and their $x$ derivatives.

Finally, the continuum limit of the string equation~\eref{stri2} splits into the system
\begin{equation}
  \label{stri2c2rcc}
  \eqalign{ \oint_{\gamma}\frac{\rmd \lambda}{2\pi\rmi}
                 V_{\lambda}(\lambda) \bV(\lambda,\bar{\epsilon};\at,\bt)
                 =
                 T_\mathrm{c}+\bar{\epsilon}^{2m} x,\cr
                 \oint_{\gamma}\frac{\rmd \lambda}{2\pi\rmi}
                 V_{\lambda}(\lambda) \bW(\lambda,\bar{\epsilon};\at,\bt)
                 =
                 T_\mathrm{c}+\bar{\epsilon}^{2m} x.}
\end{equation}
\section{Critical models with merging of two cuts  and the Painlev\'e~II hierarchy}
In this section we finally show how a particular class of solutions of the string equations for two-cut merging models
is related to the Painlev\'e~II hierarchy, but before embarking on this calculation we anticipate informally our result,
recall briefly our two-step procedure to calculate the formal asymptotic expansion of the recurrence coefficient
and discuss the source of the technical complications we will have to deal with.

Theorem~3 will essentially state that the coefficient $\at_1(x)$ in the formal asymptotic expansion of the recurrence
coefficient~\eref{a,b,c} for a symmetric solution of a two-cut merging model of order $m$
satisfies the $m$-th Painlev\'e~II equation. The general strategy to prove this theorem is again (i) to calculate large $N$
expansions for the generating functions $\bV$ and $\bW$ whose coefficients are functions of the $\at_k$ and $\bt_k$
(and their derivatives) by using the continuum limit of the ``separated odd-$n$ even-$n$'' version of theorem~1
given by equations~\eref{resc2ac}, and (ii) to substitute these expansions for $\bV$ and $\bW$ into the continuum limit
of the ``separated odd-$n$ even-$n$'' version of the string equation~\eref{stri2} given by equations~\eref{stri2c2rcc},
perform the contour integrations, and solve recursively to obtain differential equations for the coefficients $\at_k$ and
$\bt_k$ of the odd-$n$ and even-$n$ terms of the $r_{n,N}$ expansions. This recursive solution is, however, difficult
to carry out in full generality because the order $k$ in~\eref{RcS} at which the $x$ dependent term in~\eref{stri2c2rcc}
enters the expansions depends on the order of the critical model.
\subsection{Symmetric solutions of the string equations for two-cut merging models}
In this section we discuss asymptotic expansions of the recurrence coefficient~\eref{a,b,c} for which
$\bt_k(x)=(-1)^k \at_k(x)$ (symmetric solutions), and show that for these solutions the odd and even generating
functions $\bV$ and $\bW$ can be replaced by a single generating function $\mathbb{V}$.
\begin{defi}
A solution $\at(\bar{\epsilon},x)$,  $\bt(\bar{\epsilon},x)$ of the system~\eref{stri2c2rcc} is said to be symmetric if
\begin{equation}
  \label{sims}
  \bt(\bar{\epsilon},x) =\at(-\bar{\epsilon},x).
\end{equation}
\end{defi}
Note that if the functions $\bV(\lambda,\bar{\epsilon};\at(\bar{\epsilon},x),\at(-\bar{\epsilon},x))$ and
$\bW(\lambda,\bar{\epsilon};\at(\bar{\epsilon},x),\at(-\bar{\epsilon},x))$ satisfy the scaled quadratic
system~\eref{resc2ac} so do the functions
\begin{equation}
  \eqalign{
                \widetilde{\bV}(\lambda,\bar{\epsilon};\at(\bar{\epsilon},x),\at(-\bar{\epsilon},x))
                =
                \bW(\lambda,-\bar{\epsilon};\at(-\bar{\epsilon},x),\at(\bar{\epsilon},x))\cr
                \widetilde{\bW}(\lambda,\bar{\epsilon};\at(\bar{\epsilon},x),\at(-\bar{\epsilon},x),)
                =
                \bV(\lambda,-\bar{\epsilon};\at(-\bar{\epsilon},x),\at(\bar{\epsilon},x)).}
\end{equation}
Hence symmetric solutions satisfy
\begin{equation}
  \label{sims2}
  \bW(\lambda,\bar{\epsilon};\at(\bar{\epsilon},x),\at(-\bar{\epsilon},x))
  =
  \bV(\lambda,-\bar{\epsilon};\at(-\bar{\epsilon},x),\at(\bar{\epsilon},x)).
\end{equation}
Consequently,  it is convenient to introduce the function
\begin{equation}
  \label{newv}
  \mathbb{V}(\lambda,\bar{\epsilon};x)
  =
  \bV(\lambda,\bar{\epsilon};\at(\bar{\epsilon},x),\at(-\bar{\epsilon},x))
\end{equation}
since the scaled string equations~\eref{stri2c2rcc} reduce to a single equation for $\mathbb{V}$, namely
\begin{equation}
  \label{newst}
  \oint_{\gamma}\frac{\rmd \lambda}{2\pi\rmi} V_{\lambda}(\lambda)
  \mathbb{V}^{[k]}(\lambda;\at_1,\ldots,\at_k)
  =
 \delta_{k,0}T_\mathrm{c}+ \delta_{k,2m}x,\quad k\geq 0.
\end{equation}

The function $ \mathbb{V}$ has an expansion of the form
\begin{equation}
  \label{expv}
  \mathbb{V}(\lambda,\bar{\epsilon};x)
  =
  \sum_{k\geq 0} \mathbb{V}^{[k]}(\lambda;\at_1,\ldots,\at_k)\bar{\epsilon}^k,
\end{equation}
and the corresponding coefficients $\mathbb{V}^{[k]}$ are determined using the first equation of~\eref{resc2ac}
\begin{equation}
  \label{rescfin}
  \fl
  \at(\bar{\epsilon},x)(\mathbb{V}(\lambda,\bar{\epsilon};x) + \mathbb{V}(\lambda,-\bar{\epsilon};x-\bar{\epsilon}))
  (\mathbb{V}(\lambda,\bar{\epsilon};x) + \mathbb{V}(\lambda,-\bar{\epsilon};x+\bar{\epsilon}))
  =
  \lambda(\mathbb{V}(\lambda,\bar{\epsilon};x)^2-1).
\end{equation}
For example, the first few coefficients are
\begin{eqnarray}
 \mathbb{V}^{[0]} = \frac{\lambda}{w_\mathrm{c}},\\
 \mathbb{V}^{[1]} = \frac{2\at_1}{w_\mathrm{c}},\\
\label{idos}Ê \mathbb{V}^{[2]} = \frac{2\at_2}{w_\mathrm{c}} + \frac{\lambda(8r_\mathrm{c}\at_2-2\at_1^2)}{w_\mathrm{c}^{3}},\\
\label{itres}Ê  \mathbb{V}^{[3]} = \frac{2\at_3}{w_\mathrm{c}} +\frac{8r_\mathrm{c}^2\at_1''-4\at_1^3
                                                                               +\lambda(4\at_1\at_2-2r_\mathrm{c}\at_1'')}{w_\mathrm{c}^3}.
\end{eqnarray}
\subsection{Structure of the coefficients $\mathbb{V} ^{[k]}$}
To proceed further we need a general expression for the structure of the coefficients $\mathbb{V} ^{[k]}$
as functions of $\at_1,\ldots,\at_k$ and their $x$ derivatives, which could be derived by a direct analysis
of the quadratic equation~\eref{rescfin}. Although straightforward in principle, in practice the resulting
intermediate expressions are complicated. We achieve a certain simplification by forming
suitable linear combinations of shifted equations that exhibit well-defined parity. In fact, we resort to
this method twice (propositions~1 and~2).
\begin{prop}
The following linear expressions in $\mathbb{V}$  are even functions of $\bar{\epsilon}$:
\begin{eqnarray}
  \label{lea}
  \fl
   \lambda\mathbb{V}(\lambda,\bar{\epsilon};x-{\bar{\epsilon}/{2}})
   -
   \at(\bar{\epsilon},x-{\bar{\epsilon}/{2}}) \left[ \mathbb{V}(\lambda,\bar{\epsilon};x-{\bar{\epsilon}/{2}})
  +
  \mathbb{V}(\lambda,-\bar{\epsilon};x-3{\bar{\epsilon}/{2}})\right],\\
  \label{leb}
  \fl
  \lambda\mathbb{V}(\lambda,\bar{\epsilon};x+{\bar{\epsilon}/{2}})
  -
  \at(\bar{\epsilon},x+{\bar{\epsilon}/{2}}) \left[ \mathbb{V}(\lambda,\bar{\epsilon};x+{\bar{\epsilon}/{2}})
 +
 \mathbb{V}(\lambda,-\bar{\epsilon};x+3{\bar{\epsilon}/{2}}) \right].
\end{eqnarray}
\end{prop}
\noindent
\emph{Proof}. If we  perform the shift $x\rightarrow x-\bar{\epsilon}/2$ in~\eref{rescfin}  we get
\begin{eqnarray}
  \fl
  \lambda(\mathbb{V}(\bar{\epsilon};x-\bar{\epsilon}/2)^2-1)
 =
 \at(\bar{\epsilon},x-\bar{\epsilon}/2) \left[ \mathbb{V}(\bar{\epsilon};x-\bar{\epsilon}/2)
                                                          +\mathbb{V}(-\bar{\epsilon};x-3\bar{\epsilon}/2)\right]\nonumber\\
 \quad {}\times \left[\mathbb{V}(\bar{\epsilon};x-\bar{\epsilon}/2)+\mathbb{V}(-\bar{\epsilon};x+\bar{\epsilon}/2)\right].
\end{eqnarray}
The difference between this equation and its version with the substitution $\bar{\epsilon}\rightarrow -\bar{\epsilon}$ yields
\begin{eqnarray}
  \fl
 \lambda\left[\mathbb{V}(\bar{\epsilon};x-\bar{\epsilon}/2)-\mathbb{V}(-\bar{\epsilon};+\bar{\epsilon}/2)\right]
 =
 \at(\bar{\epsilon},x-\bar{\epsilon}/2)
 \left[\mathbb{V}(\bar{\epsilon};x-\bar{\epsilon}/2)  + \mathbb{V}(-\bar{\epsilon};x-3\bar{\epsilon}/2) \right]
 \nonumber\\
 \quad{} - \at(-\bar{\epsilon},x+\bar{\epsilon}/2)
                \left[ \mathbb{V}(-\bar{\epsilon};x+\bar{\epsilon}/2)
                         + \mathbb{V}(\bar{\epsilon};x+3\bar{\epsilon}/2) \right]
\end{eqnarray}
which means that~\eref{lea} holds. Similarly, if we introduce the shift $x\rightarrow x+\bar{\epsilon}/2$
in~\eref{rescfin} and perform the difference between the resulting  equation and its version with the
substitution $\bar{\epsilon}\rightarrow -\bar{\epsilon}$ we get~\eref{leb}.

The vanishing of the coefficients for odd powers of  $\bar{\epsilon}$ in~\eref{lea} and~\eref{leb} 
provide us with a series of $\lambda$-dependent constraints. To take advantage of these constraints
we first make explicit the $\lambda$ dependence of $\mathbb{V}$ in a convenient form.
Thus, using recursion in~\eref{rescfin}  we deduce that the functions $\mathbb{V}^{[k]}$ can be written as
\begin{equation}
  \label{expv1}
  \eqalign{
                \mathbb{V}^{[2i]}
                =
                \frac{1}{w_\mathrm{c}} \left( C^{[2i]} + \sum_{j=1}^{i}\left( f_j(\lambda) A_{j}^{[2i]} +g_j(\lambda)B_{j}^{[2i]}\right) \right),
                \quad  i\geq 1,\cr
                \mathbb{V}^{[2i+1]}
                =
                \frac{1}{w_\mathrm{c}} \left( C^{[2i+1]}
                + \sum_{j=1}^{i}\left( f_j(\lambda) A_{j}^{[2i+1]} +g_j(\lambda) B_{j}^{[2i+1]}\right) \right),
                \quad i\geq 1,}
\end{equation}
where
\begin{equation}
  \label{fg}
  f_j(\lambda) = \frac{(\lambda-4r_\mathrm{c})^{j+1}}{w_\mathrm{c}^{2j}}
                      = \frac{\lambda-4r_\mathrm{c}}{\lambda^j},
  \quad
  g_j(\lambda) = \frac{\lambda^{j+1}}{w_\mathrm{c}^{2j}}
                      = \frac{\lambda}{(\lambda-4r_\mathrm{c})^j}
\end{equation}
and  $A_{j}^{[k]}, B_{j}^{[k]}, C^{[k]} $ are $\lambda$-independent polynomials in  $\at_1,\ldots,\at_k$ and their
$x$ derivatives. Then $\mathbb{V}$ can be expressed in the form
\begin{equation}
  \label{vs}
  \mathbb{V} = \frac{1}{w_\mathrm{c}}
                        \left( \lambda + \mathbb{V}_0
                                + \sum_{j\geq 1} \left( f_j(\lambda)  \mathbb{V}_0^{[j]} +g_j(\lambda)  \mathbb{V}_1^{[j]} \right)\right),
\end{equation}
where
\begin{equation}
  \label{decd}
                \mathbb{V}_0 = \sum_{k\geq 1}  C^{[k]}\bar{\epsilon}^k,
                \quad
                \mathbb{V}_0^{[j]} = \sum_{i\geq 0}A_{j}^{[2j+i]}\bar{\epsilon}^{2j+i},
                \quad
                \mathbb{V}_1^{[j]} = \sum_{i\geq 0}B_{j}^{[2j+i]}\bar{\epsilon}^{2j+i}.
\end{equation}
For example, from~\eref{idos} and~\eref{itres} we find
\begin{equation}
  \label{dos}  
  \eqalign{
     A_{1}^{[2]} = 0,
  & \quad A_{1}^{[3]} = \frac{1}{2}\at_1''-\frac{\at_1^3}{4r_\mathrm{c}^2},\cr
     B_{1}^{[2]} = \frac{1}{2r_\mathrm{c}}(4r_\mathrm{c}\at_2-\at_1^2),
  & \quad B_{1}^{[3]} = \frac{\at_1}{4r_\mathrm{c}^2}(4r_\mathrm{c}\at_2-\at_1^2),\cr
     C^{[2]} = \frac{\at_1^2}{2r_\mathrm{c}},
  & \quad C^{[3]} = 2\at_3 - \frac{1}{2}\at_1''+ \frac{\at_1}{2r_\mathrm{c}^2}(\at_1^2-2r_\mathrm{c}\at_2).}
\end{equation}
\begin{prop}
The following  expressions are even functions of $\bar{\epsilon}$:
\begin{equation}
  \label{lea2}
  \fl
  \eqalign{
  \mathbb{V}_0(\bar{\epsilon};x-{\bar{\epsilon}/{2}})+ \mathbb{V}_0^{[1]}(\bar{\epsilon};x-{\bar{\epsilon}/{2}})+ \mathbb{V}_1^{[1]}
  (\bar{\epsilon};x-{\bar{\epsilon}/{2}})-2 \at(\bar{\epsilon},x-{\bar{\epsilon}/{2}}),
  \cr
  4r_\mathrm{c}\mathbb{V}_0^{[1]}(\bar{\epsilon};x-{\bar{\epsilon}/{2}})+\at(\bar{\epsilon},x-{\bar{\epsilon}/{2}})
  \Big[\mathbb{V}_0(\bar{\epsilon};x-{\bar{\epsilon}/{2}})+
  \mathbb{V}_0(-\bar{\epsilon};x-3{\bar{\epsilon}/{2}}) \Big],
  \cr
  \mathbb{V}_0^{[j+1]}(\bar{\epsilon};x-{\bar{\epsilon}/{2}})
  -\at(\bar{\epsilon},x-{\bar{\epsilon}/{2}})\Big[\mathbb{V}_0^{[j]}(\bar{\epsilon};x-{\bar{\epsilon}/{2}})+
  \mathbb{V}_0^{[j]}(-\bar{\epsilon};x-3{\bar{\epsilon}/{2}}) \Big],
  \cr
  4r_\mathrm{c}\mathbb{V}_1^{[j]}(\bar{\epsilon};x-{\bar{\epsilon}/{2}})+
  \mathbb{V}_1^{[j+1]}(\bar{\epsilon};x-{\bar{\epsilon}/{2}})
  \cr
  \quad{}-\at(\bar{\epsilon},x-{\bar{\epsilon}/{2}})\Big[\mathbb{V}_1^{[j]}(\bar{\epsilon};x-{\bar{\epsilon}/{2}})+
  \mathbb{V}_1^{[j]}(-\bar{\epsilon};x-3{\bar{\epsilon}/{2}}) \Big].}
\end{equation}
\end{prop}
\noindent
\emph{Proof}. It is enough to substitute~\eref{vs} into the expression~\eref{lea} and identify coefficients
in the $\lambda$-dependent functions $\lambda,1, f_j(\lambda),g_j(\lambda)\ (j\geq 1)$ taking into account that
\begin{eqnarray*}
  \lambda f_1=\lambda-4r_\mathrm{c},\quad \lambda g_1=\lambda+4r_\mathrm{c}g_1,\\
  \lambda f_j=f_{j-1},\quad \lambda g_j=-4r_\mathrm{c}g_j+g_{j-1},\quad j\geq 2.
\end{eqnarray*}

Equating to zero the coefficients of the odd powers of $\bar{\epsilon}$ in the third and fourth
expressions~\eref{lea2} we obtain a series of equations involving  the functions $A_{j}^{[k]}$, $B_{j}^{[k]}$  and
$\at_i(i=1,\ldots,k)$.  Moreover, since~\eref{leb} follows from~\eref{lea} under the substitutions
\begin{equation}
  \mathbb{V}(\bar{\epsilon};x)\rightarrow   \mathbb{V}(-\bar{\epsilon};x),
  \quad
  \at(\bar{\epsilon},x)\rightarrow\at(-\bar{\epsilon},x),
\end{equation}
then the relations provided by~\eref{leb} are those supplied by~\eref{lea2} with the substitutions
\begin{equation}
  \label{simm}
  A_{j}^{[k]}\rightarrow (-1)^kA_{j}^{[k]},
  \quad
  B_{j}^{[k]}\rightarrow (-1)^kB_{j}^{[k]},
  \quad \at_i\rightarrow (-1)^i \at_i.
\end{equation}

Setting to zero the coefficients of $\bar{\epsilon}^{2j+1}$ and $\bar{\epsilon}^{2j+3}$ in~\eref{lea2}
we obtain
\begin{eqnarray}
  \label{una}
  \fl
  -\at_1A_{j}^{[2j]}+ r_\mathrm{c}( A_{j}^{[2j]})' = 0,
  \\
  \label{dosb}
  \fl
  -2 \at_1 A_{j}^{[2j+2]}+A_{j+1}^{[2j+3]}
  -\at_1(A_{j}^{[2j+1]})'+2r_\mathrm{c}(A_{j}^{[2j+2]})'
  -\frac{1}{2}(A_{j+1}^{[2j+2]})' +r_\mathrm{c}( A_{j}^{[2j+1]})'' = 0,
  \\
  \label{tres}
  \fl
  -\at_1B_{j}^{[2j]}+2r_\mathrm{c} B_{j}^{[2j+1]}=0,
  \\
  \fl
  -2 \at_3 B_{j}^{[2j]}-2 \at_1B_{j}^{[2j+2]}+4r_\mathrm{c} B_{j}^{[2j+3]}+B_{j+1}^{[2j+3]}+B_{j}^{[2j]}  \at_2'
  \nonumber\\
  \fl\quad
  {}+2 \at_2  (B_{j}^{[2j]})'- \at_1'  (B_{j}^{[2j]})'-\at_1(B_{j}^{[2j+1]})' - \frac{1}{2}  (B_{j+1}^{[2j+2]})'
  \\
  \label{cuatro}
  \fl\quad
  {}-\frac{1}{4} B_{j}^{[2j]}  \at_1''-\frac{5}{4} \at_1 (B_{j}^{[2j]})''
  + \frac{3}{2}r_\mathrm{c}   (B_{j}^{[2j+1]})''+\frac{1}{2}r_\mathrm{c}(B_{j}^{[2j]})'''=0.
\end{eqnarray}
If we now sum and subtract these equations with their corresponding versions under
the substitution~\eref{simm} we get for $j\geq 1$
\begin{eqnarray}
  \label{ee1}
  A_{j}^{[2j]} = 0,\\
  \label{ee2}
  2r_\mathrm{c}\partial_x A_{j}^{[2j+2]}=\at_1\partial_x A_{j}^{[2j+1]},\\
  \label{ee3}
  A_{j+1}^{[2j+3]}=-r_\mathrm{c}\partial_x^2 A_{j}^{[2j+1]}+2\at_1A_{j}^{[2j+2]},\\
  \label{ee4}
  2r_\mathrm{c}B_{j}^{[2j+1]}=\at_1B_{j}^{[2j]},\\
  \label{ee5}
  2r_\mathrm{c}\partial_x B_{j+1}^{[2j+2]}
  =
  (2r_\mathrm{c}^2\partial_x^3+2(4r_\mathrm{c}\at_2-\at_1^2)\partial_x+(4r_\mathrm{c}\at_2-\at_1^2)_x)B_{j}^{[2j]}.
\end{eqnarray}
Furthermore, using the expression~\eref{dos} for $B_{1}^{[2]}$  we have that~\eref{ee5} can be rewritten as
\begin{equation}
  \label {ee5b}
  \partial_x B_{j+1}^{[2j+2]}
  =
  (r_\mathrm{c}\partial_x^3+2B_{1}^{[2]}\partial_x+(B_{1}^{[2]})_x)B_{j}^{[2j]}.
\end{equation}
\subsection{Symmetric solutions of the string equations and the Painlev\'e~II hierarchy}
Let us see now how  the Painlev\'e II hierarchy emerges from the string equation of two-cut merging models.
Substituting~\eref{vs}  into~\eref{newst} and identifying powers of $\bar{\epsilon}$ we obtain an infinite series
$\Sigma_i\,(i\geq 0)$  of systems of two equations. The system $\Sigma_0$ is~\eref{e0}, and the system $\Sigma_i$
for $i\geq 1$ is  given by
\begin{equation}
  \label{newst0}
  \eqalign{
                \sum_{j=1}^i(\varphi_jA_j^{[2i]}(\at_1,\ldots,\at_{2i})+\gamma_jB_j^{[2i]}(\at_1,\ldots,\at_{2i}))
                =
                \delta_{i,m}x,\cr
                \sum_{j=1}^i(\varphi_jA_j^{[2i+1]}(\at_1,\ldots,\at_{2i+1})+\gamma_jB_j^{[2i+1]}(\at_1,\ldots,\at_{2i+1}))
                =0,\quad i\geq 1,}
\end{equation}
where
\begin{equation}
  \label{ces}
  \varphi_j = \oint_{\gamma}\frac{\rmd \lambda}{2\pi\rmi}\frac{V_{\lambda}}{w_\mathrm{c}} f_j(\lambda),
  \quad
  \gamma_j = \oint_{\gamma}\frac{\rmd \lambda}{2\pi\rmi} \frac{V_{\lambda}}{w_\mathrm{c}} g_j(\lambda),  \quad  j\geq 1.
\end{equation}
Moreover, in view of the constraints~\eref{idd}--\eref{idd2} for two-cut merging singular models of order $m$ we have
\begin{equation}
  \label{iddm}
  \varphi_1 = \cdots = \varphi_{m-1} =0,
  \quad
  \varphi_m\neq 0,
  \quad
  \gamma_1\neq 0.
\end{equation}
Therefore the first $m$ systems~\eref{newst0} reduce to
\begin{equation}
  \label{newst0mm}
  \eqalign{
                \sum_{j=1}^i\gamma_jB_j^{[2i]} = 0,\cr
                \sum_{j=1}^i\gamma_jB_j^{[2i+1]} = 0,}
\end{equation}
for $ i=1,\ldots,m-1$, and
\begin{eqnarray}
  \label{newst0m}
  \eqalign{
                \varphi_mA_m^{[2m]}+\sum_{j=1}^m \gamma_jB_j^{[2m]} = x,\cr
                \varphi_mA_m^{[2m+1]}+\sum_{j=1}^m \gamma_jB_j^{[2m+1]}=0,}
\end{eqnarray}
for $i=m$. We are now ready to prove the following theorem:
\begin{theo} The coefficient $u=\at_1$ for a symmetric solution of  a two-cut merging model of order $m$
satisfies the $m$-th Painlev\'e~II equation  ($P_{II}^{m}$ equation)
  \begin{equation}
    2r_\mathrm{c} \varphi_m(r_\mathrm{c})  R_m(u)+x u=0,
  \end{equation}
where $R_m(u)$ is the differential polynomial in $u$ determined recursively  by
\begin{equation}
  \label{pan}
  \eqalign{
                R_{m+1}=-r_\mathrm{c}  \partial_{xx} R_m+2u  S_m,\cr
                2r_\mathrm{c} \partial_x S_m=u  \partial_x R_m,}
\end{equation}
with
\begin{equation}\label{in}
R_0=-\frac{1}{2r_\mathrm{c} } u_x,\quad S_0=-\frac{1}{8r_\mathrm{c}^2} u^2,
\end{equation}
and
\begin{equation}
  \varphi_m(r_\mathrm{c}) = 
  \oint_{\gamma}\frac{\rmd \lambda}{2\pi\rmi}
  \frac{(\lambda-4 r_\mathrm{c})^{m+1} V_{\lambda}}{w_\mathrm{c}^{2m+1 }}.
\end{equation}
\end{theo}
\noindent
\emph{Proof}
From~\eref{newst0mm} we see that the system $\Sigma_1$ is
\begin{equation}
  \gamma_1B_1^{[2]}=0,
  \quad
  \gamma_1B_1^{[3]}=0,
\end{equation}
so that $B_1^{[2]}=B_1^{[3]}\equiv 0$ . Then using~\eref{ee2} and~\eref{ee5b} recursively  we get
\begin{equation}
  B_j^{[2j]}= B_j^{[2j+1]}\equiv 0, \quad j\geq 1.
\end{equation}
Equating to zero the coefficients of $\bar{\epsilon}^{2j+5}$ and $\bar{\epsilon}^{2j+7}$ in the fourth expression
of~\eref{lea2} we obtain
\begin{eqnarray}
  \label{ee4b}
  2r_\mathrm{c}B_{j}^{[2j+3]}=\at_1B_{j}^{[2j+2]},\\
  \label{ee5c}
  \partial_x B_{j+1}^{[2j+4]}
  =
  (-r_\mathrm{c}\partial_x^3+2B_{1}^{[2]}\partial_x+(B_{1}^{[2]})_x) B_{j}^{[2j+2]}.
\end{eqnarray}
Then, since the system $\Sigma_2$ is
\begin{equation}
  \gamma_1B_1^{[4]}=0,
  \quad
  \gamma_1B_1^{[5]}=0,
\end{equation}
it follows that $B_1^{[4]}=B_1^{[5]}\equiv 0$. Hence, using~\eref{ee4b} and~\eref{ee5c}
recursively  we get
\begin{equation}
  B_j^{[2j+2]}=B_j^{[2j+3]}\equiv 0, \quad j\geq 1.
\end{equation}
Repeating this process using the systems $\Sigma_i$ up to $i=m-1$ we get
\begin{equation}
  B_j^{[2m]}= B_j^{[2m+1]}=0,\quad j=2,\ldots,m,
\end{equation}
and
\begin{equation}
  \label{fin1}
  2r_\mathrm{c}B_{1}^{[2m+1]}=\at_1B_{1}^{[2m]}.
\end{equation}
Moreover, taking into account~\eref{ee1}, the system $\Sigma_m$ reduces to
\begin{equation}
  \label{finm}
  \eqalign{
                \gamma_1B_1^{[2m]}=x,\cr
                \varphi_mA_m^{[2m+1]}+ \gamma_1B_1^{[2m+1]}=0.}
\end{equation}
Therefore, using~\eref{fin1} we get
\begin{equation}
  \label{pain1}
  2r_\mathrm{c}\varphi_mA_m^{[2m+1]}(\at_1)+\at_1x=0,
\end{equation}
which in view of~\eref{ee2} and~\eref{ee3} proves that $\at_1$ is a solution of the $(m-1)$-th
member of the Painlev\'e~II hierarchy~\eref{pan} with
\begin{equation}
  R_m(u)=A_m^{[2m+1]}(u),\quad S_m=A_m^{[2m+2]}(u).
 \end{equation}
Moreover, taking into account~\eref{pan}  and  the expression of  $A_1^{[3]}(u)$ in~\eref{aes}
we get~\eref{in}.

Again, a more precise notation for $R_m(u)$ and $\varphi_m(r_\mathrm{c})$ in theorem~3 would be $R_m(r_\mathrm{c},u)$
and $\varphi_m(r_\mathrm{c},\mathbf{g})$, which emphasizes the explicit dependence on the critical value $r_\mathrm{c}$
and the coupling constants $\mathbf{g}$.
\subsection{The quartic model and the $P_{II}^{1}$ equation}
Although the derivation of theorem~3 has been somewhat complicated, its applications are fairly simple.
For example, we have already seen that the quartic model~\eref{bip}  at $T_\mathrm{c}=g_2^2/(4g_4)$ is a
two-cut merging model of order $m=1$ with $r_\mathrm{c}=-g_2/(4g_4)$. It also follows that $\varphi_1=2g_2$.
Thus we have that the first coefficient $u = \at_1$ of a symmetric solution~\eref{sims} of the corresponding string
equation in the double scaling limit  satisfies the $P_{II}^{1}$ equation~\cite{BL03}
\begin{equation}\label{firstPII}
  g_2^2 u_{xx} - 2g_4 (4g_4u^3+xu) = 0.
\end{equation}
\section{Concluding remarks}
In this paper we have presented a method to characterize and compute the large $N$ formal asymptotics of regular and critical
Hermitian matrix models with general even potentials in the one-cut and two-cut cases. This method also leads to an explicit
formulation, in terms of coupling constants and critical parameters, of the members of the Painlev\'e~I and Painlev\'e~II hierarchies associated with one-cut and two-cut critical models respectively.

As we pointed out in the introduction, the asymptotic form of the recurrence coefficient in the multi-cut case in general is
not represented by (integer or fractional) power series in $N^{-1}$ but involves a quasi-periodic dependence on $N$.
Asymptotic expansions of the recurrence coefficients containing explicitly this quasi-periodic  dependence on $N$
have been considered by Bleher and Eynard~\cite{BL03b} to prove the presence of the  Painlev\'e~II hierarchy in
a class of nonsymmetric models with merging of two cuts.  It would be interesting to investigate if our method can be generalized
to deal with this type of asymptotic expansions. This generalization would require new appropriate formulations of the continuum
limits for both the recurrence coefficients $r_{n,N}$ and $s_{n,N}$, as well as for the generating function $U_{n,N}$.
A possible application could be the investigation of critical models  featuring the birth of a cut~\cite{EY06,CL08},
which do not seem to correspond to any conformal field theory and it is not clear if any integrable equation underlies
the asymptotics of the associated recurrence coefficients in the double scaling limit.
\ack
The financial support of the Universidad Complutense under project GR35/10-A910556, the Comisi\'on Interministerial de
Ciencia y Tecnolog\'{\i}a under projects FIS2008-00200 and FIS2008-00209 are gratefully acknowledged.

\end{document}